\documentclass[preprint]{aastex}

\usepackage{amsmath}
\usepackage{bm}

\IfFileExists{vc.tex}{\input{vc.tex}}{\newcommand{\githash}{794a221}\newcommand{\giturl}{https://www.github.com/AnnieJumpCannon/rave}}

\newcommand{\acronym}[1]{{\small{#1}}}

\newcommand{\project}[1]{\textsl{#1}}
\newcommand{\gaia}{\project{Gaia}}
\newcommand{\thecannon}{\project{The~Cannon}}
\newcommand{\rave}{\project{\acronym{RAVE}}}
\newcommand{\galah}{\project{\acronym{GALAH}}}
\newcommand{\ges}{\project{Gaia-ESO}}
\newcommand{\apogee}{\project{\acronym{APOGEE}}}
\newcommand{\aspcap}{\project{\acronym{ASPCAP}}}
\newcommand{\lamost}{\project{\acronym{LAMOST}}}
\newcommand{\hipparcos}{\project{Hipparcos}}
\newcommand{\epic}{\project{K2/EPIC}}

\newcommand{\tgas}{\project{\acronym{TGAS}}}
\newcommand{\raveon}{\rave-on}

\newcommand{\teff}{T_{\mathrm{eff}}}
\newcommand{\logg}{\log g}
\newcommand{\feh}{[\mathrm{Fe/H}]}

\newcommand{\Nspectra}{520,781}
\newcommand{\Nstars}{457,588}

\newcommand{\ReportedStellarParameters}{441,397} 
\newcommand{\ReportedAbundances}{1,685,851} 

\newcommand{\Dvector}[1]{\boldsymbol{#1}}
\newcommand{\vectheta}{\Dvector{\theta}}
\newcommand{\vecv}{\Dvector{v}}
\newcommand{\argmin}[1]{\underset{#1}{\operatorname{argmin}}\,}

\begin{document}
\title{The \raveon\ catalog of stellar atmospheric parameters and\\
chemical abundances for chemo-dynamic studies in the \gaia\ era}

\author{
Andrew~R.~Casey\altaffilmark{1},
Keith~Hawkins\altaffilmark{1},
David~W.~Hogg\altaffilmark{2,3,4,5},
Melissa~Ness\altaffilmark{5},
Hans~Walter-Rix\altaffilmark{5},
Georges~Kordopatis\altaffilmark{6},
Andrea~Kunder\altaffilmark{6},
Matthias~Steinmetz\altaffilmark{6},
Sergey~Koposov\altaffilmark{1},
Harry~Enke\altaffilmark{6},
Jason~Sanders\altaffilmark{1},
Gerry~Gilmore\altaffilmark{1},
Toma\v{z}~Zwitter\altaffilmark{6},
Kenneth~C.~Freeman\altaffilmark{7},
Luca~Casagrande\altaffilmark{7},
Gal~Matijevi\v{c}\altaffilmark{6},
George~Seabroke\altaffilmark{8},
Olivier~Bienaym\'e\altaffilmark{9},
Joss~Bland-Hawthorn\altaffilmark{10},
Brad~K.~Gibson\altaffilmark{11},
Eva~K.~Grebel\altaffilmark{12},
Amina~Helmi\altaffilmark{13},
Ulisse~Munari\altaffilmark{14},
Julio~F.~Navarro\altaffilmark{15,16},
Warren~Reid\altaffilmark{17,18},
Arnaud~Siebert\altaffilmark{9},
Rosemary~Wyse\altaffilmark{19}\\
\vspace{2em}
\emph{(Affiliations can be found at the end of the article)}
}

\begin{abstract}
The orbits, atmospheric parameters, chemical abundances, and ages of 
individual stars in the Milky Way provide the most comprehensive 
illustration of galaxy formation available.  The Tycho-\gaia\ 
Astrometric Solution (\tgas) will deliver astrometric parameters for
the largest ever sample of Milky Way stars, though its full potential
cannot be realized without the addition of complementary spectroscopy.  
Among existing spectroscopic surveys, the RAdial Velocity Experiment 
(\rave) has the largest overlap with \tgas\ ($\gtrsim$200,000 stars).  
We present a data-driven re-analysis of \Nspectra\ \rave\ spectra using 
\thecannon.  For red giants, we build our model using high-fidelity 
\apogee\ stellar parameters and abundances for stars that overlap 
with \rave.  For main-sequence and sub-giant stars, our model uses 
stellar parameters from the \epic.  We derive and validate 
effective temperature $\teff$, surface gravity $\logg$, and chemical 
abundances of up to seven elements (O, Mg, Al, Si, Ca, Fe, Ni).  We 
report a total of \ReportedAbundances\ elemental abundances with a 
typical precision of 0.07~dex, a substantial improvement over previous 
\rave\ data releases.  The synthesis of \raveon\ and \tgas\ is the 
most powerful data set for chemo-dynamic analyses of the Milky Way ever produced.
\end{abstract}


\section{Introduction} 
\label{sec:introduction}

The Milky Way is considered to be our best laboratory for understanding galaxy
formation and evolution.  This premise hinges on the ability to precisely measure 
the astrometry and chemistry for (many) individual stars, and to use those data 
to infer the structure, kinematics, and chemical enrichment of the Galaxy 
\citep[e.g.,][]{Nordstrom_2004,Schlaufman_2009,Deason_2011,Casagrande_2011,
Ness_2012,Ness_2013a,Ness_2013b,Casey_2012,Casey_2013,Casey_2014a,Casey_2014b,
Boeche_2013,Kordopatis_2015,Bovy_2016}.  However, these quantities are not known
for even 1\% of stars in the Milky Way.  Stellar distances are famously imprecise 
\citep[e.g.,][]{van_Leeuwen_2007,Jofre_2015,Madler_2016}, proper motions can be
plagued by unquantified systematics from the first epoch observations 
\citep[e.g.,][]{Casey_Schlaufman_2015}, and stellar spectroscopists frequently 
report significantly different chemical abundance patterns from the same spectrum 
\citep{Smiljanic_2014}.  The impact these issues have on scientific inferences 
cannot be understated.  Imperfect astrometry or chemistry limits understanding 
in a number of sub-fields in astrophysics, including the properties of exoplanet 
host stars, the formation (and destruction) of star clusters, as well as studies
of stellar populations and Galactic structure, to name a few.

The \gaia\ mission represents a critical step forward in understanding the Galaxy.
\gaia\ is primarily an astrometric mission, and will provide precise positions,
parallaxes and proper motions for more than $10^9$ stars in its final data
release in 2022.  While this is a sample size about four orders of magnitude 
larger than its predecessor \hipparcos, both astrometry and chemistry are 
required to fully characterize the formation and evolution of the Milky Way. 
\gaia\ will also provide radial velocities, stellar parameters and chemical 
abundances for a subset of brighter stars, but these measurements will not be 
available in the first few data releases. Until those abundances are available,
astronomers seeking to simultaneously use chemical and dynamical information are
reliant on ground-based spectroscopic surveys to complement the available 
\gaia\ astrometry.

The first \gaia\ data release will include the Tycho-Gaia Astrometric Solution
\citep[hereafter \tgas;][]{Michalik_2015a,Michalik_2015b}: positions, proper 
motions, and parallaxes for approximately two million stars in the Tycho-2 
\citep{Hog_2000} catalog.  After cross-matching all major stellar spectroscopic 
surveys\footnote{Specifically we cross-matched the Tycho-2 catalog against the 
\apogee\ DR13 \citep{Zasowski_2013}, \ges\ internal DR4 
\citep{Gilmore_2012,Randich_2013}, \galah\ internal DR1 \citep{DeSilva_2015}, 
\lamost\ DR1 \citep{Cui_2012}, and \rave\ DR4 \citep{Steinmetz_2006} catalogs.}, 
we found that the RAdial Velocity Experiment \citep[\rave;][]{Steinmetz_2006} 
survey is expected to have the largest overlap with the first \gaia\ 
data release: up to 264,276 stars.  We used the \gaia\ universe model snapshot 
\citep{Robin_2012} to estimate the precision in parallax and proper motions that
could be available in the first \gaia\ data release (DR1) for stars in those 
overlap samples.  Comparing the expected precision to what is currently available, 
we further found that the \rave\ survey will benefit most from \gaia\ DR1: the
distances of 63\% of stars in the \rave--\gaia\ DR1 overlap sample are expected 
to improve with the first \gaia\ data release, and 47\% of stars are likely to
have better proper motions.  Although the \gaia\ universe model assumes 
end-of-mission uncertainties --- and does not account for systematics in the 
first data release --- this calculation still provides intuition for the relative
improvement that the first \gaia\ data release can make to ground-based surveys.  
The expected improvements for \rave\ motivated us to examine what chemical 
abundances were available from those data, and to evaluate whether we could 
enable new chemo-dynamic studies by contributing to the existing set of chemical
abundances.

We briefly describe the \rave\ data in Section~\ref{sec:data}, before explaining
our methods in Section~\ref{sec:method}.  In Section~\ref{sec:validation}
we outline a number of validation experiments, including: internal sanity checks,
comparisons with literature samples, and investigations to ensure our results
are consistent with expectations from astrophysics.  We discuss the implications
of these comparisons in Section~\ref{sec:discussion}, and conclude with instructions
on how to access our results electronically.

\section{Data}
\label{sec:data}

\rave\ is a magnitude-limited stellar spectroscopic survey of the (nearby) Milky Way,
principally designed to measure radial velocities for up to $10^6$ stars.
Observations were conducted on the 1.2~m UK Schmidt telescope at the Australian 
Astronomical Observatory\footnote{Formerly the Anglo-Australian Observatory.} from 
2003--2013.  A large 5.7~degree field-of-view and robotic fibre positioner made for 
very efficient observing:  spectra for up to 150 targets could be simultaneously
acquired.  When observations concluded in April 2013, at least \Nspectra\ useful 
spectra had been collected of more than \Nstars\ unique objects.

The target selection for \rave\ is based on the $I$-band apparent magnitude,
$9 < I < 12$, with a weak $J - K_s > 0.5$ cut near the disk and bulge \citep{Wojno_2016}.  
The $I$ band was used for the target selection because it has a good overlap with the
wavelength range that \rave\ operates in:  8410--8795~\AA.  The resolution and 
wavelength coverage of \rave\ is comparable to the Radial Velocity Spectrometer on
board the \gaia\ space telescope \citep{Munari_2005,Kordopatis_2011,Recio-Blanco_2016}, 
and the wavelength range overlaps with one of the key setups used for the ground-based 
high-resolution \ges\ survey \citep{Gilmore_2012,Randich_2013}.  The spectral region 
includes the \ion{Ca}{2} near infrared triplet lines --- strong transitions that 
are dominated by pressure broadening --- which are visible even in metal-poor stars
or spectra with very low signal-to-noise (S/N) ratios.  Atomic transitions of 
light-, $\alpha$-, and Fe-peak elements are also present, allowing for detailed 
chemical abundance studies.

The exposure times for \rave\ observations were optimised to obtain radial 
velocities for as many stars as possible.  Detailed chemical abundances were
always an important science goal of the survey, but this was a secondary objective.  
For this reason the distribution of S/N ratios in \rave\ spectra is considerably 
lower than other stellar spectroscopic surveys where chemical abundances are the 
primary motivation.  The \rave\ spectra have an effective resolution 
$\mathcal{R} \approx 7{,}500$ and the distribution of S/N ratios peaks at 
$\approx$50~pixel$^{-1}$.  For comparison, the \galah\ survey 
\citep{DeSilva_2015} --- which was specifically constructed for detailed chemical 
abundance analyses --- includes a wavelength range about 2.5 times larger at 
resolution $\mathcal{R} \approx 28{,}000$, and yet the \galah\ project still 
targets for S/N $\gtrsim100$ per resolution element.

Despite the relatively low resolution and S/N of the spectra compared to other
surveys, the \rave\ data releases have provided excellent radial velocities, 
stellar atmospheric parameters ($\teff$, $\logg$), and detailed chemical abundances
\citep{Steinmetz_2006,Zwitter_2008,Siebert_2011,Boeche_2011,Kordopatis_2013,
Kunder_2016}.  In this work we make use of spectra that has been reprocessed for 
the fifth \rave\ data release.  These re-processing steps include: a detailed 
re-reduction of all the original data frames, with flux variances propagated at 
every step; an updated continuum-normalization procedure; as well as revised 
determinations of stellar radial velocities and morphological classifications. 
At the end of this processing for each survey observation we were provided with:
rest-frame wavelengths (without resampling), continuum-normalized fluxes, $1\sigma$
uncertainties in the continuum-normalized flux values, as well as relevant metadata
for each observation.  We refer the reader to the official fifth data release paper
of the \rave\ survey, as presented by \citet{Kunder_2016}, for more details of this
re-processing.

Given the high-quality of the normalization performed by the \rave\ team, we chose
not to re-normalize the spectra.  Our tests demonstrated that the procedure 
outlined in \citet{Kunder_2016} is sufficient for our analysis procedure. Therefore,
there were a limited number of pre-processing steps that we performed before starting
our analysis.  First, we calculated inverse variance arrays from the $1\sigma$ 
uncertainties provided, and then we re-sampled the flux and inverse variance
arrays onto a common rest-wavelength map for all stars.  Depending on the fibre 
used and the stellar radial velocity, the range of rest-frame wavelength values
varied for each star.  Given that fluxes were unavailable in the edge pixels for 
most stars, we excluded pixels outside of the rest wavelength range 
$8423.2\,{\rm \AA} \le \lambda \le 8777.6\,{\rm \AA}$.  This corresponds to about
30~pixels excluded on either side of the common wavelength array, leaving us with
945~pixels per spectrum for science.

\section{Method}
\label{sec:method}

We chose to adopt a data-driven model for this analysis, in contrast to the
physics-based models used in \rave\ data releases to date.  Specifically, we will
use an implementation of \thecannon\ \citep{Ness_2015,Ness_2016}.  Although this 
choice complicated the construction of our model (e.g., see Section 
\ref{sec:the-training-set}), a data-driven approach makes use of all available 
information in the spectrum and lowers the S/N ratio at which systematic effects 
begin to dominate.  In other words, in the low S/N regime, a well-constructed 
data-driven model will yield more precise \emph{labels} (e.g., stellar parameters 
and chemical abundances) than most physics-driven models\footnote{However, see 
\citet{Casey_2016a}.}.  This is particularly relevant for the low-resolution 
\rave\ data analysed here, because about half of the spectra have S/N 
$\lesssim 50$~pixel$^{-1}$.

There are two main analysis steps when using \thecannon: the \emph{training} 
step and the \emph{test} step.  We describe these stages in the context of our
model in the following section, and a more thorough introduction can be found
in \citet{Ness_2015}.  We make the following explicit assumptions about the 
\rave\ spectra and \thecannon:

\begin{itemize}
\item We assume that any fibre- and time-dependent variations in spectral
resolution in the \rave\ spectra are negligible.
\item The \rave\ noise variances are approximately correct, independent between
pixels, and normally distributed.
\item We assume that the normalization procedure employed by the \rave\ pipeline
is invariant with respect to the labels we seek to measure (e.g., $\teff$, $\logg$,
or [Fe/H]), and invariant with respect to the S/N ratio.  In other words, we assume
that the normalization procedure does not produce different results for high S/N
spectra compared to low S/N spectra, nor does the normalization procedure vary 
non-linearly with respect to stellar parameters (e.g., [Fe/H]).
\item We assume that stars with similar labels ($\teff$, $\logg$, and abundances)
have similar spectra.
\item A stellar spectrum is a smooth function of the label values for that star,
and we assume that the function is smooth enough within a sub-space of the labels
(e.g., the giant branch or the main-sequence) that it can be reasonably approximated 
with a low-order polynomial in label space.
\item The training set (Section~\ref{sec:the-training-set}) has mean accurate labels
for most, but not all stars. That is to say that we do not assume that \emph{every} 
label in the training is accurate. We can afford to have a small fraction of 
inaccurate labels; a few obvious misclassifications in the training set are affordable.
\item We assume that the training data are similar (in spectra) to the test data 
where they overlap in label space, and we assume that the training data spans enough
of the label space to capture the variation in the test-set spectra.
\end{itemize}

\subsection{The model}
\label{sec:the-model}

\noindent{}Given our assumptions, the model we adopt is
\begin{eqnarray}\label{eq:model}
y_{jn} & = & \vecv(\ell_n)\cdot\vectheta_j + e_{jn}\quad ,
\end{eqnarray}

\noindent{}where $y_{jn}$ is the pseudo-continuum-normalized flux for star $n$ at wavelength pixel
$j$, $\vecv(\ell_n)$ is the vectorizing function that takes as input the $K$ labels
$\ell_n$ for star $n$ and outputs functions of those labels as a vector of length
$D>K$, $\vectheta_j$ is a vector of length $D$ of parameters influencing the model at
wavelength pixel $j$, and $e_{jn}$ is the residual (noise).  Here we will only consider
vectorizing functions with second-order polynomial expansions (e.g., $\teff^2$, see Sections 
\ref{sec:a-simple-model}--\ref{sec:evolved-star-model}).  The noise values $e_{jn}$ can 
be considered to be drawn from a Gaussian distribution with zero mean and variance 
$\sigma_{jn}^2 + s_j^2$, where $\sigma_{jn}^2$ is the variance in flux $y_{jn}$ and 
$s_j^2$ describes the excess variance at the $j$-th wavelength pixel.

At the \emph{training} step we fix the $K$-lists of labels for the $n$ training set stars.
At each wavelength pixel $j$, we then find the parameters $\vectheta_j$ and $s_j^2$
by optimizing the penalized likelihood function
\begin{eqnarray}\label{eq:train}
\vectheta_j,s^2_j &\leftarrow& \argmin{\vectheta,s}\left[
    \sum_{n=0}^{N-1} \frac{[y_{jn}-\vecv(\ell_n)\cdot\vectheta]^2}{\sigma^2_{jn}+s^2}
    + \sum_{n=0}^{N-1} \ln(\sigma^2_{jn}+s^2) + \Lambda{}\,Q(\vectheta)
    \right]
  \quad ,
\end{eqnarray}

\noindent{}where $\Lambda$ is a regularization parameter which we will heuristically set
in later sections, and $Q(\vectheta)$ is a L1 regularizing function that encourages 
$\vectheta$ values to take on zero values without breaking convexity \citep{Casey_2016b}:
\begin{eqnarray}\label{eq:regularization-function}
	Q(\vectheta) = \sum_{d=1}^{D-1} |{\theta_d}| \quad .
\end{eqnarray}

Note that the $d$ subscript here is zero-indexed; the function $Q(\vectheta)$ does not act
on the (first) $\theta_0$ coefficient, as this is a `pivot point' (mean flux value) that 
we do not expect to diminish with increasing regularization (e.g., see equation 
\ref{eq:vectorizer-three-label}).  In practice we first fix $s_j^2 = 0$ to make equation
\ref{eq:train} a convex optimization problem, then we optimize for $\vectheta_j$, before 
solving for $s_j^2$.

The \emph{test step} is where we fix the parameters $\vectheta_j,s_j^2$ at all wavelength
pixels $j$, and optimize the $K$-list of labels $\ell_m$ for the $m$-th test set star.  Here
the objective function is:
\begin{eqnarray}\label{eq:test}
  \ell_m &\leftarrow& \argmin{\ell}\left[
    \sum_{j=0}^{J-1} \frac{[y_{jm}-\vecv(\ell)\cdot\vectheta_j]^2}{\sigma_{jm}^2 + s_j^2}
    \right]
  \quad .
\end{eqnarray}

After optimizing equation \ref{eq:test} for the $m$-th star we store the covariance matrix 
$\bm{\Sigma}_m$ for the labels $\ell_m$, which provides us with the formal errors on $\ell_m$. 
The formal errors are expected to be underestimated, and in Section~\ref{sec:validation} 
we judge the veracity of these errors through validation experiments.

\subsection{The training set}
\label{sec:the-training-set}

We sought to construct a training set of stars across the main-sequence, the
sub-giant branch, and the red giant branch.  We required stars with precisely measured
effective temperature $\teff$, surface gravity $\logg$, and elemental abundances
of O, Mg, Si, Ca, Al, Fe, and Ni.  This proved to be difficult because the magnitude
range of \rave\ does not overlap substantially with high-resolution spectroscopic
surveys.  The fourth internal data release of the \ges\ survey includes 
giant and main-sequence stars, but only 142 overlap with \rave, which is too small to
be a useful training set for our purposes.  The thirteenth data release from the 
\project{Sloan Digital Sky Survey} \citep{sloan_dr13} includes labels for \apogee\ stars on the
giant branch and (uncalibrated values for) the main-sequence, but our tests indicated
that the \apogee\ main-sequence labels suffered from significant systematic effects.  
A flat, then `up-turning' main-sequence is present, and the metallicity gradient trends in 
the opposite direction with respect to $\logg$ on the main-sequence (i.e., metal-poor
stars incorrectly sit above an isochrone in a classical Hertzsprung-Russell diagram).
If we consider lower-resolution studies as potential training sets, there are 2,369
stars that overlap with \lamost\ --- of which 2,213 have positive S/N ratios in the 
$g$-band (\texttt{snrg}).  However, the labels are expectedly less precise given the
lower resolution, there are no elemental abundances available for the main-sequence 
stars\footnote{Abundance information is available for \lamost\ stars from \citet{Ho_2016},
but that sample contains only giant stars.}, and the \lamost\ lower main-sequence suffers
from the same systematic effects seen in the \apogee\ data.

These constraints forced us to construct a heterogeneous training set.  Given previous
successes in transferring high S/N ratio labels from \apogee\ \citep{Ness_2015,
Ness_2016,Ho_2016,Casey_2016b}, we chose to use the 1,355 stars in the \apogee---\rave\ 
overlap sample for giant star labels in the training set.  Of these, about 900 are 
giants according to \apogee.  From this sample we selected stars to have: 
determinations in all abundances of interest ([X/H] $> -5$ for O, Mg, Al, Si, Ca, Fe, 
and Ni); S/N ratios of $>$200~pixel$^{-1}$ in \apogee\ and $>$25~pixel$^{-1}$ in \rave; 
and we further required that the \aspcap\ did not report any peculiar flags 
(\texttt{ASPCAPFLAG = 0}).  These restrictions left us with 536 stars along the giant 
branch, with metallicities ranging from $[{\rm Fe/H}] = -1.79$ to 0.26.  Intermediate 
tests with globular cluster members showed that the metallicity range of the training 
set needed to extend at least below $[{\rm Fe/H}] \lesssim -2$ in order for our catalog 
to be practically useful.  Without additional metal-poor stars, the lowest metallicity
labels reported by our model would be around $[{\rm Fe/H}] \approx -2$, even for well
studied stars with $[{\rm Fe/H}] \sim -4$ (e.g., CD~38-245).  For this reason we
supplemented our sample of \apogee\ giant stars with 176 known metal-poor giant stars 
observed by \rave.  The effective temperature $\teff$, surface gravity $\logg$ and
iron abundance [Fe/H] labels were adopted from \citet{Fulbright_2010, Ruchti_2011}.
For this sample of metal-poor stars, we assumed that the elemental abundances of O, 
Mg, Al, Si, Ca, and Ni followed typical trends of Galactic chemical evolution: we
asserted $[{\rm Mg/Fe}] = +0.4$, $[{\rm O/Fe}] = +0.4$, $[{\rm Al/Fe}] = -0.5$, 
$[{\rm Ca/Fe}] = +0.4$, $[{\rm Si/Fe}] = +0.4$, and $[{\rm Ni/Fe}] = -0.25$.  We stress
that this decision is made solely to ensure that our overall metallicity scale reflects
that of the \rave\ survey, down to $[{\rm Fe/H}] \sim -4$.  Indeed, it is likely that 
for most of these elements, these abundances cannot be measured from \rave\ spectra for 
ultra metal-poor stars: the atomic transitions in the \rave\ spectral region are simply
too weak to influence the spectrum.  For this reason, our adopted abundances for these
very metal-poor stars represent an `anchor point' in order to ensure our overall 
metallicity scale is correct.  We do not recommend the use of our individual abundance
labels at $[{\rm Fe/H}] \sim -4$.  We discuss this issue in more detail in Section~\ref{sec:discussion}.

Assembling a suitable training set for the main-sequence and sub-giant branch was less
trivial.  There are no spectroscopic studies that extend the range of stellar types we 
are interested in (e.g., FGKM-type stars), and which also have a large enough sample size 
that overlaps with \rave.  Moreover, most of the spectroscopic studies we considered also 
showed a flat lower main-sequence, a systematic consequence of the analysis method adopted 
\citep[see][for discussion on this issue]{Bensby_2014}.  For these reasons we chose to make 
use of the \epic\ catalog \citep{Huber_2016} for the training set labels on the 
main-sequence and sub-giant branch.  The \epic\ catalog follows from the successful
\project{Kepler} input catalog \citep{Brown_2011}, and provides probabilistic stellar 
classifications for 138,600 stars in the \project{K2} fields based on the 
astrometric, asteroseismic, photometric, and spectroscopic information available for
every star.  There are 4,611 stars that overlap between \epic\ and \rave.

\epic\ differs from the \project{Kepler} input catalog because \epic\ does not 
benefit from having narrow-band $DDO_{51}$ photometry to aid dwarf/giant 
classification.  Despite this limitation, the labels in the \epic\ catalog have 
already been shown to be accurate and trustworthy \citep{Huber_2016}.  However, 
when the posteriors are wide (i.e., the quoted confidence intervals are large) 
due to limited information available, it is possible that a star has been 
misclassified.  This is most prevalent for sub-giants, where \citet{Huber_2016} 
note that $\approx55-70$\% of sub-giants are misclassified as dwarfs.  The 
probability of misclassification is usually quantified in the uncertainties given
for each star; most dwarfs that have a higher possibility of being sub-giants have
large confidence intervals.  Therefore, requiring low uncertainties will decrease 
the total sample size, but in practice it removes most misclassifications.  The 
situation is far more favourable for dwarfs and giants.  Only 1--4\% of giant 
stars are misclassified as dwarfs, and about 7\% of dwarfs are misclassified as 
giants.  To summarise, the \epic\ labels with narrow confidence intervals are 
usually of high fidelity, and given that we have spectra, we can identify any
obvious misclassifications.

\begin{figure}[p]
\includegraphics[width=\textwidth]{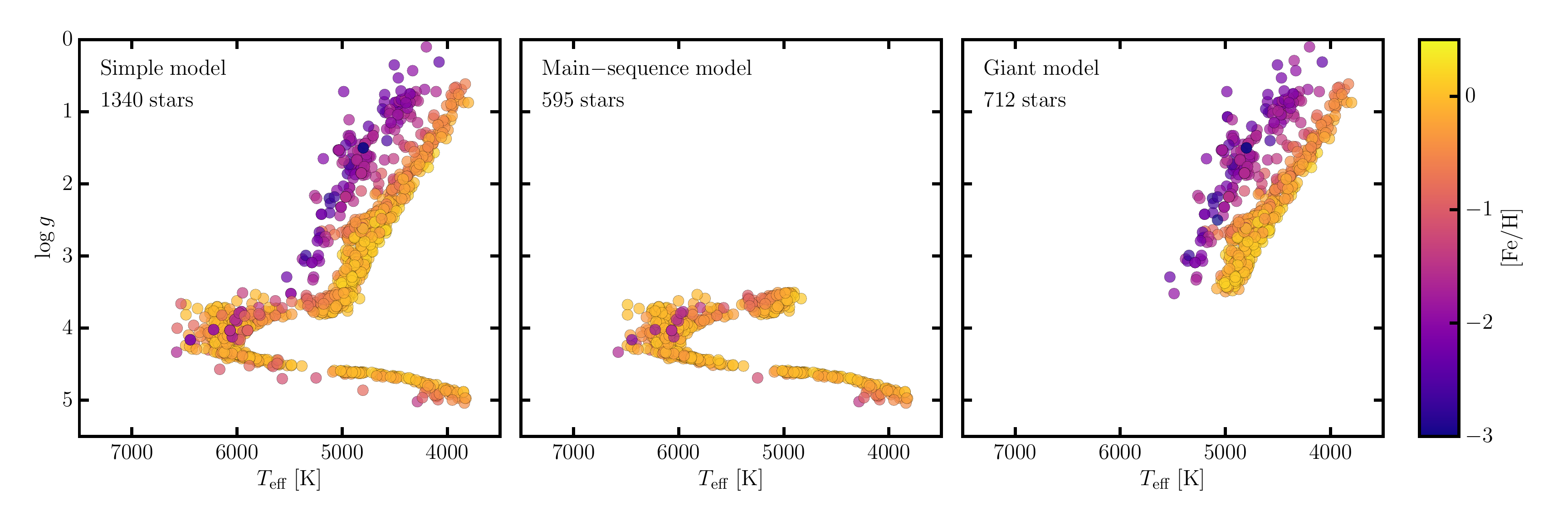}
\caption{Effective temperature $\teff$ and surface gravity $\logg$ for all stars in the training sets. Stars are colored by their metallicity [Fe/H], and the three panels show stars in the simple model (left panel; Section~\ref{sec:a-simple-model}), the main-sequence star model (middle panel; Section~\ref{sec:unevolved-star-model}), and the giant star model (right panel; Section~\ref{sec:evolved-star-model}).\label{fig:training-set-hrd}}
\end{figure}

We sought to have a small overlap between our giant and main-sequence star training
sets.  Most of our giant training set is encapsulated within $0 < \logg < 3.5$, 
however there is a sparse sampling of stars reaching to $\logg \approx 4$.  We
required $\logg > 3.5$ for the \epic\ main-sequence/sub-giant star training set,
allowing for $\approx0.5$~dex of overlap between the two training sets.  We further
employed the following quality constraints on the \epic\ catalog: the upper and lower 
confidence intervals in $\teff$ must be below 150~K; the upper and lower confidence 
intervals in $\logg$ must be less than 0.15~dex; the S/N of the \rave\ spectra must
exceed 30~pixel$^{-1}$; and $\teff \leqslant 6750$~K.  Unfortunately these strict 
constraints removed most metal-poor stars, which we later found to cause the test 
labels to have under-predicted abundances for dwarfs of low metallicity.  For this 
reason we relaxed (ignored) those quality constraints for stars with 
$[{\rm Fe/H}] < -1$, and included an additional 12 turn-off stars with 
$-1.6 \gtrsim [{\rm Fe/H}] \gtrsim -2.1$ from \citet{Ruchti_2011}.  After training
a model based on main-sequence and giant stars (Section \ref{sec:the-model}), we found 
we could identify misclassifications by leave-one-out cross-validation.  However, we 
chose not to do this because the number of likely misclassifications in the training
set was negligible ($\approx$1\%), and the improvement in main-sequence test set labels
was minimal.  The distilled sample of the \rave--\epic\ overlap catalog contains 595 
stars (583 of 4,611 from \epic).  The full training set for each model (see next 
sections) is shown in Figure~\ref{fig:training-set-hrd}.

\subsection{The simple model: a 3-label model ($\teff$, $\logg$, $\feh$) for all stars}
\label{sec:a-simple-model}

We have constructed a justified training set for stars across the main-sequence, sub-giant,
and red giant branch.  However the lack of overlap between \rave\ and other works have
resulted in a somewhat peculiar situation.  Detailed abundances are available from \apogee\
for all giant stars in our sample, however only imprecise (but accurate on expectation)  
metallicities are available from \epic\ for stars on the main-sequence and the sub-giant 
branch.  Here we will construct a simple model for \emph{all} stars that only makes use 
of three labels ($\teff$, $\logg$, [Fe/H]), before we outline how we derive abundances 
for giant branch stars.  The complexity for this model will be quadratic ($\teff^2$
is the highest term), where the vectorizer $\vecv(\ell_n)$ expands as,
\begin{eqnarray}\label{eq:vectorizer-three-label}
\vecv(\ell_n) \rightarrow \left[1, T_{{\rm eff},n}, \logg_n, [{\rm Fe/H}]_n, T_{{\rm eff},n}^2, \logg_n\,T_{{\rm eff},n}, \feh_n\,T_{{\rm eff},n}, \logg_n^2, \feh_n\,\logg_n, \feh_n^2\right]
\end{eqnarray}

\noindent{}such that $\vecv(\ell)$ produces the design matrix:
\begin{eqnarray}
	\vecv(\ell) \rightarrow \begin{bmatrix} \vecv(\ell_0) \\ \vdots \\ \vecv(\ell_{N-1}) \end{bmatrix} \quad .
\end{eqnarray}

\begin{figure*}[p]
\includegraphics[width=\textwidth]{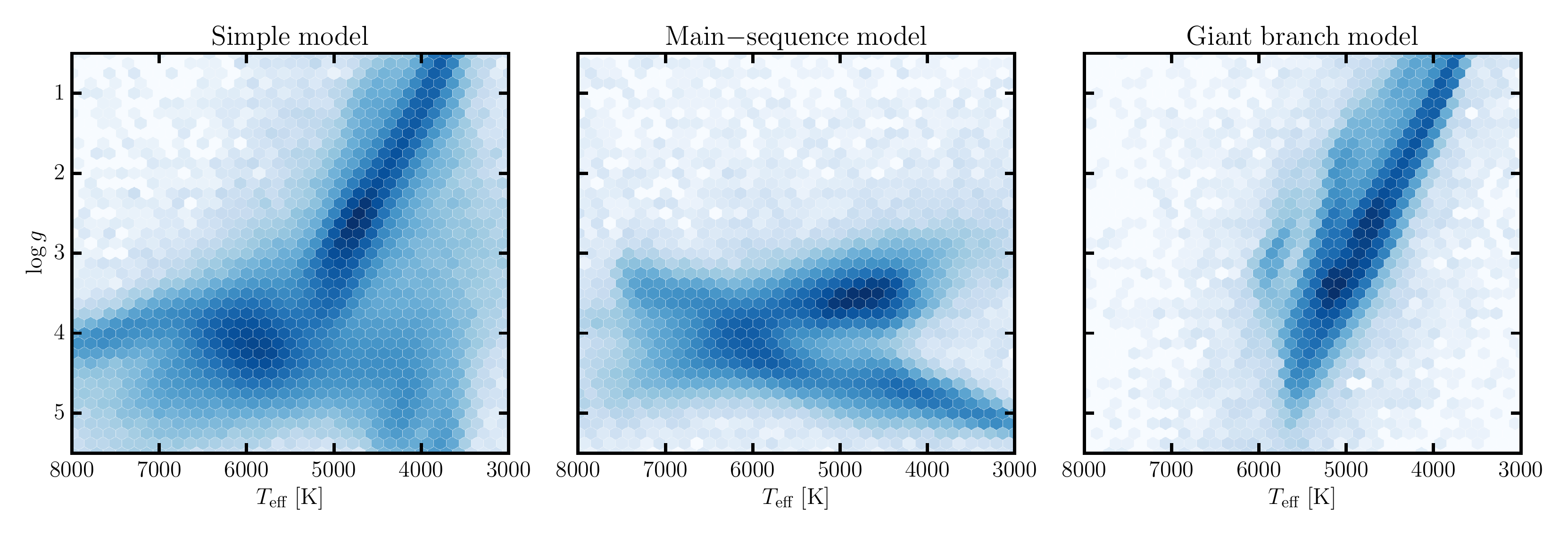}
\caption{The logarithmic density of effective temperature $\teff$ and surface gravity $\logg$ for all \Nspectra\ \rave\ spectra, as derived using the simple model (left panel; Section~\ref{sec:a-simple-model}), the main-sequence star model (center panel; Section~\ref{sec:unevolved-star-model}), and the giant star model (right panel; Section~\ref{sec:evolved-star-model}). These panels demonstrate how a single quadratic model is insufficient for all \rave\ stars (left panel), and illustrate some of the systematic artefacts that can result from testing on stars outside of the training set (center and right panel).  These panels do not represent our final results, which are shown in Figure~\ref{fig:test-set-hrd}.\label{fig:test-set-density}}
\end{figure*}

We used no regularization ($\Lambda = 0$) for this model.  After training the model we
treated all \Nspectra\ spectra as test set objects.  In the left-hand panel of Figure 
\ref{fig:test-set-density} we show the effective temperature $\teff$ and surface gravity 
$\logg$ for all spectra.  The main-sequence and red giant branch are clearly visible.  
However, the details of stellar evolution are no longer present: the sub-giant branch is 
not discernible, and there are a number of systematic artefacts (over-densities) present
in label space.  These artefacts disappear when we require additional quality constraints 
(e.g., no peculiar morphological classifications), but the complexity of the 
Hertzsprung-Russell diagram is still not present.  Thus, we concluded that while this 
model could be useful for deriving stellar classifications (e.g., F2-type giant), the 
labels are too imprecise.

We chose to adopt separate models for the main-sequence and the red giant branch rather
than switch to a single model with higher complexity.  This choice allowed us to derive
stellar parameters for stars on the main-sequence and sub-giant branch, as well as 
detailed elemental abundances for red giant branch stars.  However, adopting two separate
models introduces the challenge of how to combine the results from two models, or how to
assign one star as `belonging' to a single model.  In Section~\ref{sec:joining-the-models}
we describe how we will use the simple model introduced in this section to discriminate
between results from a 3-label main-sequence model in Section \ref{sec:unevolved-star-model} 
and a 9-label giant star model in Section \ref{sec:evolved-star-model}.

\subsection{A 3-label model ($\teff$, $\logg$, $\feh$) for unevolved stars}
\label{sec:unevolved-star-model}

We constructed a three-label quadratic model using only main-sequence and sub-giant
stars. In order to set the regularization hyperparameter $\Lambda$ for this model, we 
trained 30 models with different regularization strengths, spaced evenly in logarithmic
steps between $\Lambda = 10^{-3}$ to $\Lambda = 10^{3}$.  We then performed leave-one-out 
cross-validation for each model.  Specifically, for each star in the training set: we 
removed the star; trained the model; and then inferred labels from the removed star as 
if it was a test object.  We also performed leave-one-out cross-validation on an 
unregularized ($\Lambda = 0$) model, which we will use as the basis for comparison.  
For the unregularized case, we calculated the bias and root-mean-square (RMS) deviation 
between: the training set labels, and the labels we derived by cross-validation, where 
one star was removed at a time and the model was re-trained.  We repeated this calculation
of bias and RMS deviation for all 30 models with different regularization strengths $\Lambda$.

\begin{figure}[p]
\center
\includegraphics[width=0.45\textwidth]{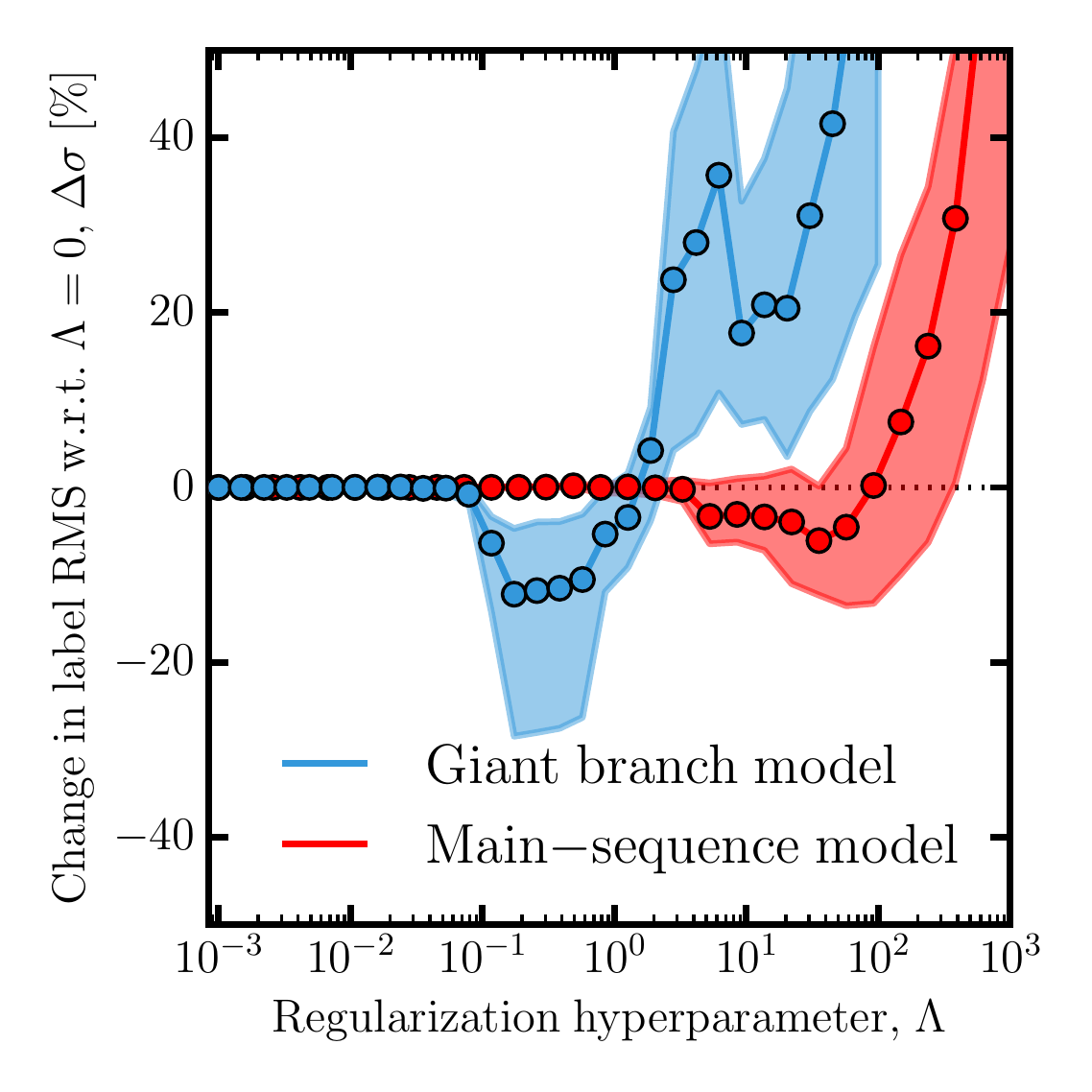}
\caption{The percentage change in RMS deviation between inferred and training labels at different regularization strengths.  The RMS values were calculated by leave-one-out cross-validation, and are shown with respect to an unregularized model ($\Lambda = 0$).  The points and solid line indicate the mean improvement across all labels. The filled area represents the minimum and maximum improvements over all labels. With increasing regularization strength, there is a minimum in the RMS deviation over all labels, which is where we set $\Lambda$ for each model (see text for details).\label{fig:set-hyperparameters}}
\end{figure}

We show the \emph{percentage difference} in the RMS deviation of the labels with respect
to the unregularized model in Figure \ref{fig:set-hyperparameters}.  The upper 
and lower envelope represent the boundaries across all labels, showing that with increasing
regularization, the RMS decreased in \emph{all} labels.  We found similar
improvements in the biases, however these were already minimal in the unregularized
case.  The improvement in RMS reaches a minimum value near 
$\Lambda = 35.6$ ($\approx10^{1.5}$), where
we achieve RMS deviations that are about 10\% better than the unregularized case.
Based on this improvement we set $\Lambda = 35.6$ for this model.  At this regularization 
strength, the bias and RMS values found by leave-one-out
cross validation are, respectively: $38$~K and $256$~K for $\teff$, $0.05$~dex and 
$0.29$~dex for $\logg$, with $0.03$~dex and $0.17$~dex for [Fe/H].

We inferred labels for all \Nspectra\ \rave\ spectra using this main-sequence/sub-giant
star model; we made no \emph{a priori} decisions as to whether a star was likely a
main-sequence/sub-giant star or not.  The results for the entire survey sample are shown
in the center panel of Figure \ref{fig:test-set-density}.  The increased density of 
solar-type stars is consistent with \rave\ observing stars in the local neighbourhood, 
and the high number of turn-off and main-sequence stars relative to the sub-giant branch 
is expected from the relative lifetimes of these evolutionary phases.  An over-density
of stars near the base of the giant branch is also present.  This artefact is due 
to having giant stars in the test set, but not in the training set, and the model is 
(poorly) extrapolating outside the convex hull of the training set.

\subsection{A 9-label model for detailed abundances of giant stars}
\label{sec:evolved-star-model}

The red giant branch stars in our training set have stellar parameters 
($\teff$, $\logg$) and up to 15 elemental abundances from the \aspcap\ 
\citep{Garcia_Perez_2016}.  A subset of these elements have atomic transitions in the 
\rave\ wavelength region: \ion{O}{1}, \ion{Mg}{1}, \ion{Al}{1}, \ion{Si}{1}, 
\ion{Ca}{2}, \ion{Ti}{1}, \ion{Fe}{1}, and \ion{Ni}{1}.  However, we 
excluded [Ti/H] from our abundance list because of systematics in the \aspcap\
[Ti/H] abundances \citep{Holtzman_2015,Hawkins_2016}.  Therefore we are left 
with nine labels in our giant star model: $\teff$, $\logg$, and seven elemental 
abundances.

Similar to Sections \ref{sec:a-simple-model} and \ref{sec:unevolved-star-model},
we used a quadratic vectorizer for the giant star model.  Here the terms are 
expanded in the same way as equation \ref{eq:vectorizer-three-label}, only
with nine labels instead of three.  We set the regularization hyperparameter $\Lambda$
in the same way described in Section \ref{sec:unevolved-star-model}, using the same 30 trials of $\Lambda$.
The results are shown in Figure \ref{fig:set-hyperparameters}, where
again the enveloped region represents the minimum and maximum change in RMS label
deviation with respect to the unregularized case.  At the point of maximum improvement 
near $\Lambda = 0.13$ $(\approx10^{-0.9})$, the RMS in all nine labels has decreased by up to $30$\%,
with all labels showing an improvement $>5$\%, and the mean improvement over all labels is about 10\%.
Near $\Lambda \approx 10^{-0.9}$ to $10^{-0.3}$, the regularization also produces a sparser matrix
of $\vectheta$, with $\approx20$\% more terms (mostly cross-terms) having zero-valued entries.
Based on the increased model sparsity and decreasing RMS deviation in the labels, 
we adopt $\Lambda = 0.57$ ($10^{-0.25}$) for the giant star model.  The bias in 
labels from a regularized model with $\Lambda = 0.57$ is negligible: $-0.3$~K in 
$\teff$, and $<$0.007~dex in magnitude for $\logg$ and all seven elemental
abundances.  The RMS at this regularization strength is 69~K in $\teff$, 0.18~dex 
in $\logg$, and varies between $0.07-0.09$~dex depending on the elemental abundance.

We inferred labels for all \Nspectra\ \rave\ spectra using this model, again without
regard for whether a star was likely a giant or not.  The results for all survey stars
are summarized in the right panel of Figure \ref{fig:test-set-density}.  The red clump
is clearly visible and in the expected location, without requiring any post-analysis 
calibration.  However, artefacts due to dwarf stars being present in the test set,
and not in the training set, are also present.

\subsection{Deriving joint estimates from multiple models}
\label{sec:joining-the-models}

We have derived labels for all \Nspectra\ \rave\ spectra using the three models
described in previous sections.  The results from our first model 
(Section~\ref{sec:a-simple-model}) --- which includes the main-sequence, sub-giant 
and red giant branch --- shows that a single 3-label quadratic model is too simple 
for the \rave\ spectral range.   The other models have problems, too: unrealistic 
over-densities in label space show that the main-sequence model and the giant model
make very poor extrapolations for stars outside their respective training sets.  
For these reasons we were forced to exclude or severely penalize incorrect results
from both models.  We emphasize that the choices here are entirely heuristic, 
and depart from interpreting \thecannon\ output labels as the maxima of individual 
likelihood functions.  Each model produces estimates of the labels for a given star, 
and we use those estimates to produce a unified estimate, but this joint estimate 
is calculated by disregarding the probabilistic attributes of individual estimates.

Before attempting to join the results from the models in Sections 
\ref{sec:unevolved-star-model} and \ref{sec:evolved-star-model}, we excluded results
in either model that had a reduced $\chi_{r}^2 > 3$.  We further discarded stars with
labels that are outside the extent of the training set.  Specifically for the
results from the giant model we (conservatively) excluded stars with derived 
$\logg > 3.5$, and for the results from the main-sequence model we excluded 
sub-giant stars ($\logg < 4$ and $\teff < 5000$~K) that were outside the two-dimensional 
($\teff$, $\logg$) convex hull of the training set used for the main-sequence model.  
Unfortunately these restrictions did not remove all spurious results.  The reason 
for this can be explained with an example:  consider that our giant star model was 
trained with only giant stars but tested with both giant stars and dwarf stars.  
Some classes of stars (e.g., metal-poor dwarfs) can project into a region of label 
space that would suggest it is a giant (e.g., a clump star).  These objects could 
have relatively low $\chi_{r}^2$ values (e.g., $\chi_{r}^2 < 3$) and in this example,
they would appear as bonafide red clump stars.  These incorrect projections are 
extrapolation errors in high dimensions that project to `normal' parts of the label 
space in two dimensions.  For these reasons we also made use of the simple model in 
Section~\ref{sec:a-simple-model} to inform whether we should adopt results from: 
the red giant branch model; the main-sequence/sub-giant model; or a linear combination 
of the two.

\begin{figure}[p]
\center
\includegraphics[width=0.5\textwidth]{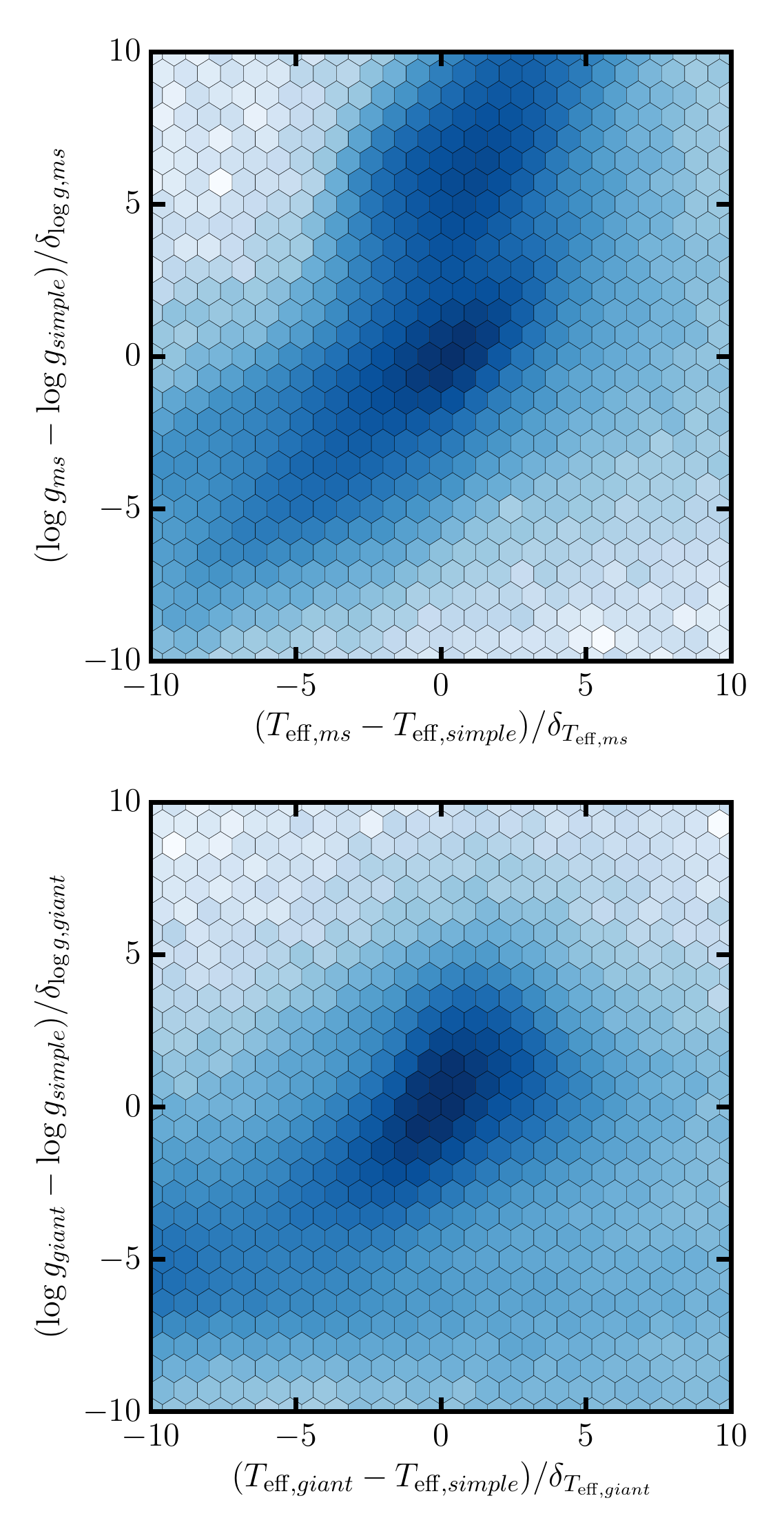}
\caption{The normalized differences in effective temperature $\teff$ and surface gravity $\logg$ between the main-sequence model and the simple model (top panel), and the giant model and the simple model (bottom panel).  The density scaling is logarithmic, and the differences in $\teff$ and $\logg$ are scaled to make them approximately isotropic (see text for details).  The peak at $(0, 0)$ represents good agreement between the simple model and comparison model, whereas the over-densities elsewhere are a consequence of testing the model on stars very different to the training set (e.g., dwarf stars tested on a model trained with only giant stars).\label{fig:joint-model-differences}}
\end{figure}

In Figure \ref{fig:joint-model-differences} we show the differences in effective 
temperature $\teff$ and surface gravity $\logg$ between: the main-sequence model 
(Section~\ref{sec:unevolved-star-model}) and the simple model (Section 
\ref{sec:a-simple-model}); and the differences between the red giant branch model 
(Section~\ref{sec:evolved-star-model}) and the simple model (Section 
\ref{sec:a-simple-model}).  We have scaled the differences in $\teff$ and $\logg$ 
to make the central peak near $(0, 0)$ to be approximately isotropic by setting
	$\delta_{T_{\rm eff}} = 90$~K for the main-sequence model, 
	$\delta_{T_{\rm eff}} = 50$~K for the giant model, and 
	$\delta_{\log{g}} = 0.15$~dex for both models.
It is important to note that these $\delta$ values do not represent any kind of 
intrinsic uncertainty or precision: they are merely normalization factors.  Adopting
substantially different scaling factors would produce in clear inconsistencies in our
results (e.g., sub-giants being misclassified as giants).  Therefore we chose these 
factors empirically to make the distributions in Figure \ref{fig:joint-model-differences}
approximately isotropic, and to some extent, comparable.  In Figure 
\ref{fig:joint-model-differences} the stars within the peak at $(0, 0)$ represent 
objects where the simple model and the comparison model both report similar labels.  
The artefacts seen in the Hertzsprung-Russell diagrams in Figure \ref{fig:test-set-density} 
are also present in Figure \ref{fig:joint-model-differences} as over-densities far away
from the central peak.  Therefore, we can adopt the scaled distance in labels 
$\teff$ and $\logg$ from the simple model to the main-sequence model $d_{ms}$,
\begin{eqnarray}
	d_{ms} & = & \left(\frac{T_{{\rm eff},ms} - T_{{\rm eff},simple}}{\delta_{T_{{\rm eff},ms}}}\right)^2 + \left(\frac{\log{g}_{ms} - \log{g}_{simple}}{\delta_{\log{g},ms}}\right)^2 \quad ,
\end{eqnarray}

\noindent{}and the scaled distance from the simple model to the giant model $d_{giant}$,
\begin{eqnarray}
	d_{giant} & = & \left(\frac{T_{{\rm eff},giant} - T_{{\rm eff},simple}}{\delta_{T_{{\rm eff},giant}}}\right)^2 + \left(\frac{\log{g}_{giant} - \log{g}_{simple}}{\delta_{\log{g},giant}}\right)^2  \quad ,
\label{eq:d_giant}
\end{eqnarray}

\noindent{}to derive the weights,
\begin{eqnarray}
	w_{ms} = \frac{1}{{d_{ms}}^2} & \text{and} & w_{giant} = \frac{1}{{d_{giant}}^2} \quad ,
\end{eqnarray}

\noindent{}and produce the weighted labels $\hat\ell$:
\begin{eqnarray}
	\hat\ell = \frac{w_{ms}\,\ell_{ms} + w_{giant}\,\ell_{giant}}{w_{ms} + w_{giant}} \quad .
\end{eqnarray}

\begin{figure}[p]
\center
\includegraphics[width=0.5\textwidth]{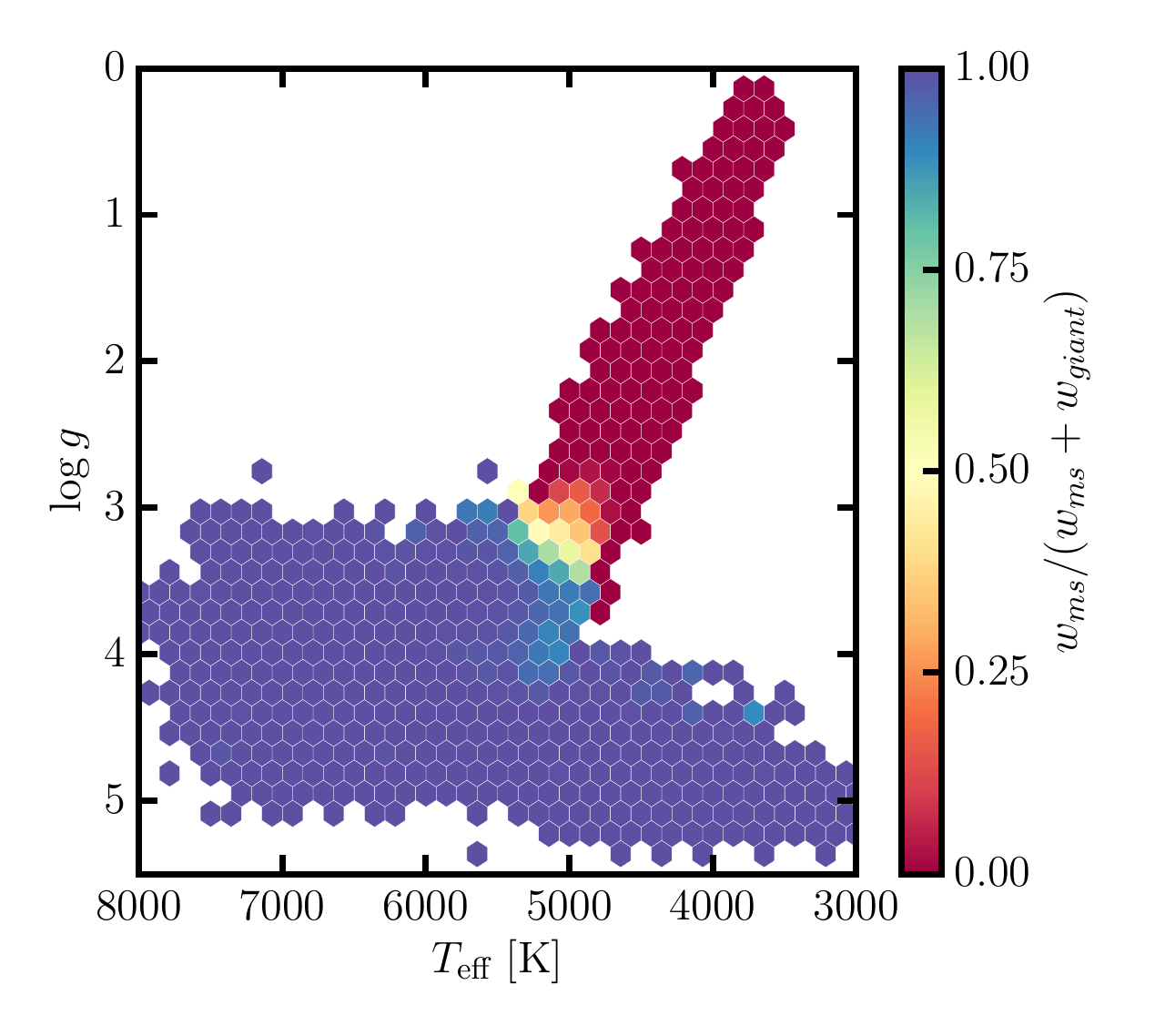}
\caption{The mean relative main-sequence model weight $w_{ms}/(w_{ms} + w_{giant})$ at each hexagonal bin of weighted effective temperature $\teff$ and surface gravity $\logg$.  The relative weighting illustrates how only results from the main-sequence model are adopted for unevolved stars, and there is a gradual transition to using results from the giant model, before only results from the giant model are used for evolved stars.\label{fig:model-weights}}
\end{figure}

We calculate weighted errors of $\hat\ell$ in the same manner.  
In Figure \ref{fig:model-weights} we show the mean relative
weight $w_{ms}/(w_{ms} + w_{giant})$ within each two-dimensional hexagonal bin of 
$\hat\teff$ and $\hat\logg$.  Hereafter when we refer to labels (e.g., $\teff$),
we refer to those from the joint estimate $\hat\ell$, not individual estimates 
from separate models.  For giant stars the relative weight of 
the main-sequence model is zero, and vice-versa for main-sequence stars.
The relative weights smoothly transition from 0 to 1 on the sub-giant branch
near $\log{g} \approx 3.5$, in the training set overlap region of both models.
For abundance labels in the giant model that are not in the main-sequence
model (e.g., [O/H], [Mg/H]), we only report abundances for objects if
$w_{ms} < 0.05$.

\begin{figure}[p]
\includegraphics[width=\textwidth]{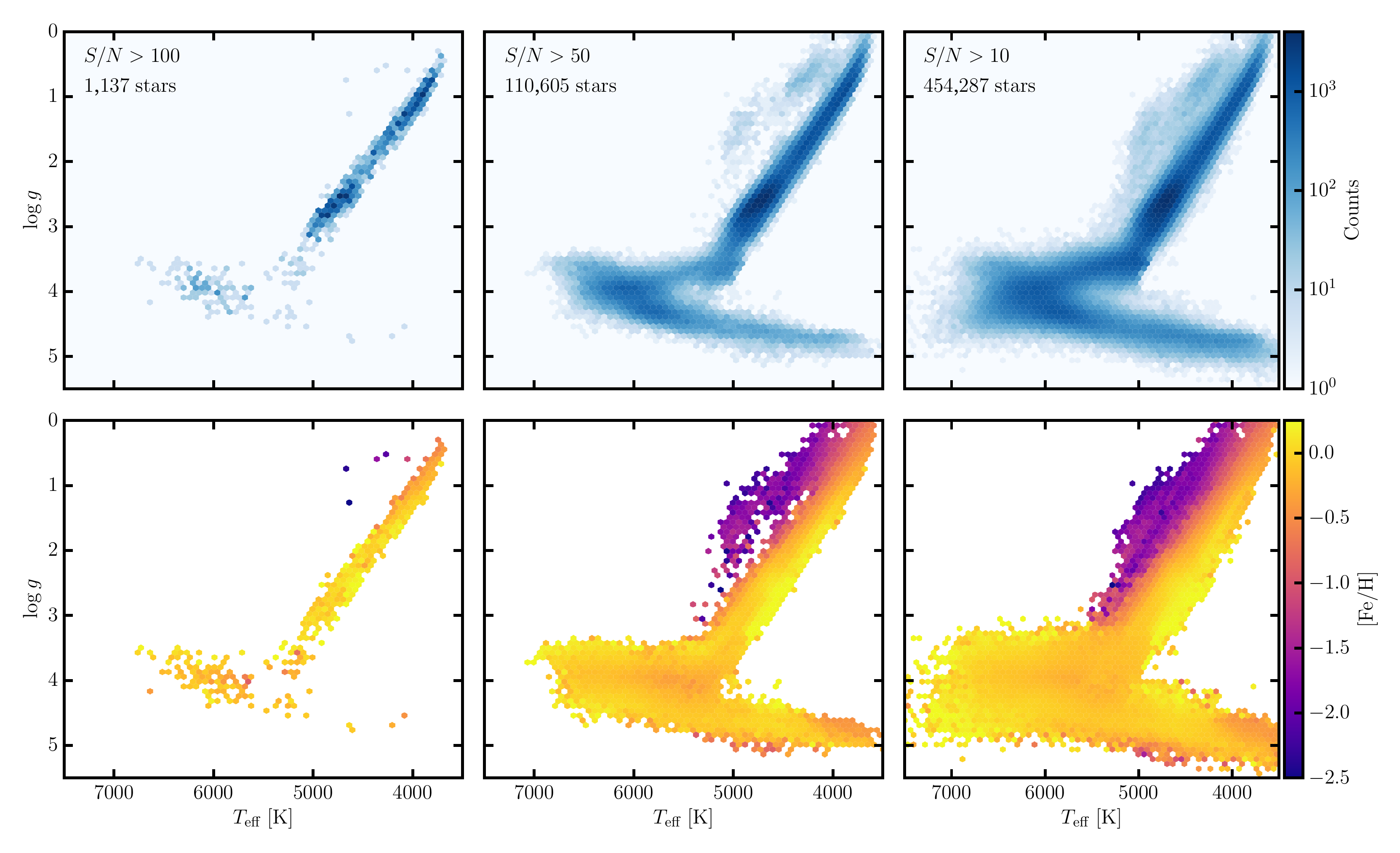}
\caption{The effective temperature $\teff$ and surface gravity $\logg$ for \rave\ stars after combining labels from the main-sequence and giant star models.  Only results meeting our quality constraints are shown (see Section~\ref{sec:discussion}). The top three panels show logarithmic density, and bins in the bottom three panels are colored by the median metallicity in each bin.\label{fig:test-set-hrd}}
\end{figure}

The weighted $\teff$ and $\logg$ values for stars meeting different S/N 
constraints are shown in Figure \ref{fig:test-set-hrd}, both in logarithmic 
density and mean metallicity.  The artefacts from individual models are no 
longer apparent, and the complete structure of the Hertzsprung-Russell diagram
is visible.  However, there are a number of caveats introduced by the decisions
we have made on how to combine estimates from multiple models. We discuss these 
issues in detail in Section \ref{sec:discussion}.

\section{Validation experiments}
\label{sec:validation}

In addition to the cross-validation tests that we have previously described, 
we have conducted a number of internal and external validation experiments to 
test the validity of our results.  We will begin by describing internal validation
tests based on repeat observations, before evaluating our accuracy based on
high-resolution literature comparisons.

\subsection{Internal validation}
\subsubsection{Repeat observations}
\label{sec:repeat-observations}

The \rave\ survey performed repeat observations for 43,918 stars with time 
intervals ranging from a few hours to up to four years.  This timing was 
constructed to be quasi-logarithmic such that spectroscopic binaries could
be optimally identified. Most of the stars that were observed multiple times
were only observed twice, with thirteen visits being the maximum number 
of observations for any target.  These repeat observations allow us to 
quantify the level of (in)correctness in our formal errors.

We calculated all possible pair-wise differences between the labels we
derived from multiple visits.  If \rave\ observed a star $H$ times, there 
are $H!/2(H-2)!$ possible $a$-to-$b$ pair-wise combinations where we can calculate 
the difference between the derived label (for example, $\logg$) over the
quadrature sum of their formal errors $(\logg_a - \logg_b)/\sqrt{{\sigma_{\logg,a}}^2 + {\sigma_{\logg,b}}^2}$.  If our derived labels were unbiased and our formal
errors were correct, the distribution of these pair-wise comparisons would be
well-represented by a Gaussian distribution with zero mean and variance
of unity.  However, our formal errors are likely to be under-estimated,
and therefore we introduce a systematic error floor for each label, which
is added in quadrature to every observation such that (for example, $\logg$),
\begin{eqnarray}
	\eta_{\logg} & = & \frac{\logg_a - \logg_b}{\sqrt{{\sigma_{\logg,a}}^2 + {\sigma_{\logg,b}}^2 + 2{\sigma_{\logg,floor}}^2}} \quad .
\end{eqnarray}

\begin{figure*}[p]
\includegraphics[width=\textwidth]{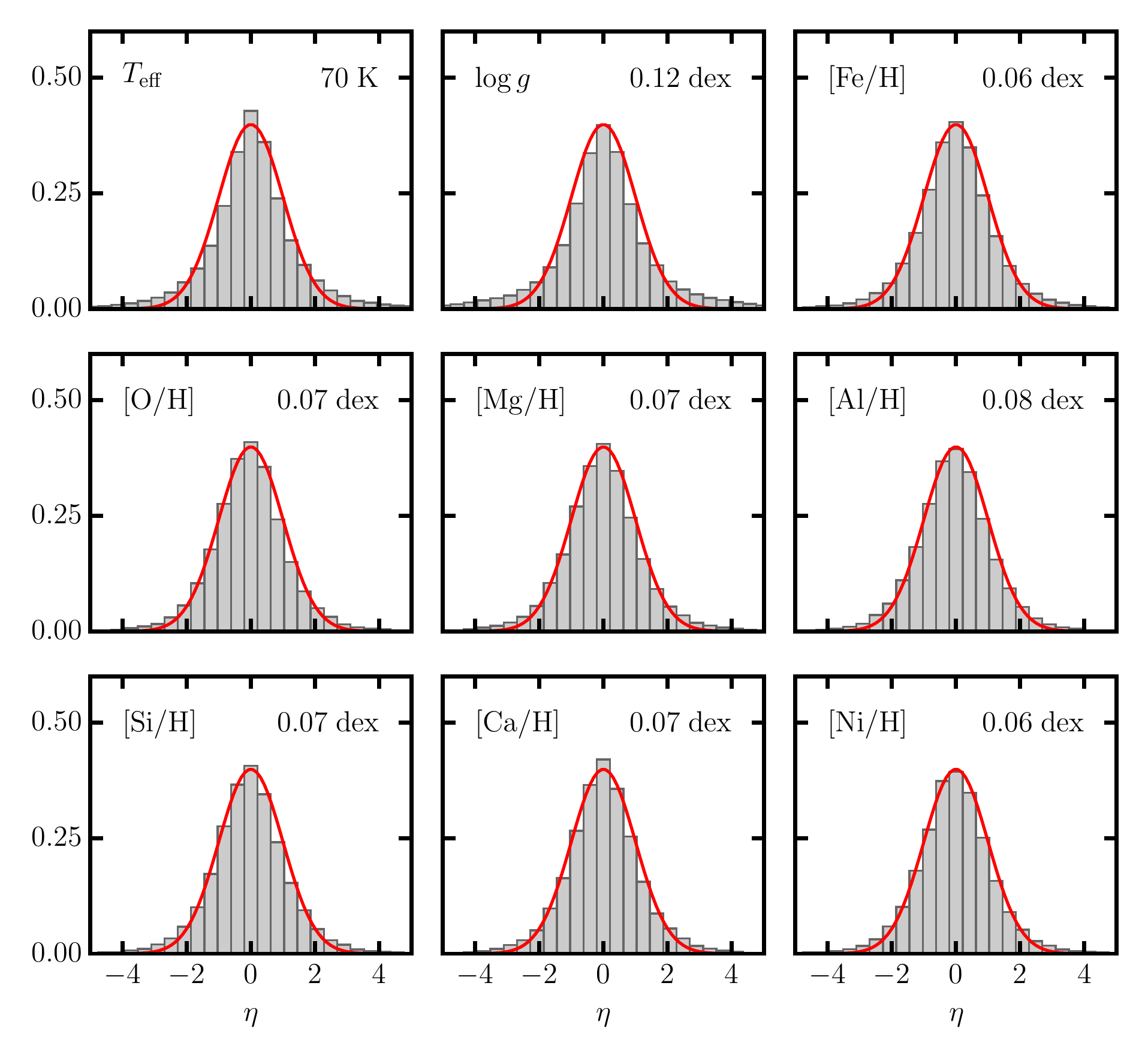}
\caption{Distribution of differences in label estimates from multiple visits, divided by the quadrature sum of their formal errors, and a minimum error value for each observation.  If our measurements were unbiased and our errors were representative, no minimum error floor would be required and the distribution of $\eta$ would by normally-distributed with zero mean and unity variance. 
We increased the error floor for each label until the variance in the distribution of $\eta$ approximately reached unity. Derived error floors are shown for each label.\label{fig:pairwise-comparison}}
\end{figure*}

We increased the minimum label error until the \emph{variance} of the $\eta$ 
distribution approximately reached unity.  We found the minimum error in
$\teff$ to be 70~K, 0.12~dex in $\logg$, and varied between 0.06--0.08~dex
for individual elements.  The minimum errors are given with the 
distributions of $\eta$ for each label in Figure \ref{fig:pairwise-comparison}.
These minimum values form part of our error model, such that they \emph{have} 
been added in quadrature with the formal errors; the quoted label errors 
in our catalog include these minimum errors.

\subsubsection{Precision as a function of S/N ratio}
\label{sec:precision-wrt-snr}

We further used the repeat observations in \rave\ to build intuition for
the label precision that was achievable as a function of S/N ratio.  
Specifically we stacked all spectra for a given star by summing the 
fluxes weighted by the inverse variances, then treated the stacked spectra
as normal survey stars.  We inferred labels for the stacked spectra 
using all three models, and derived a joint estimate as per Section 
\ref{sec:joining-the-models}.  The labels we inferred from each stacked 
spectrum then served as a basis of comparison for the labels we derived 
from the individual visit spectra of the same star, which are of lower 
S/N ratios.

\begin{figure*}[p]
\includegraphics[width=\textwidth]{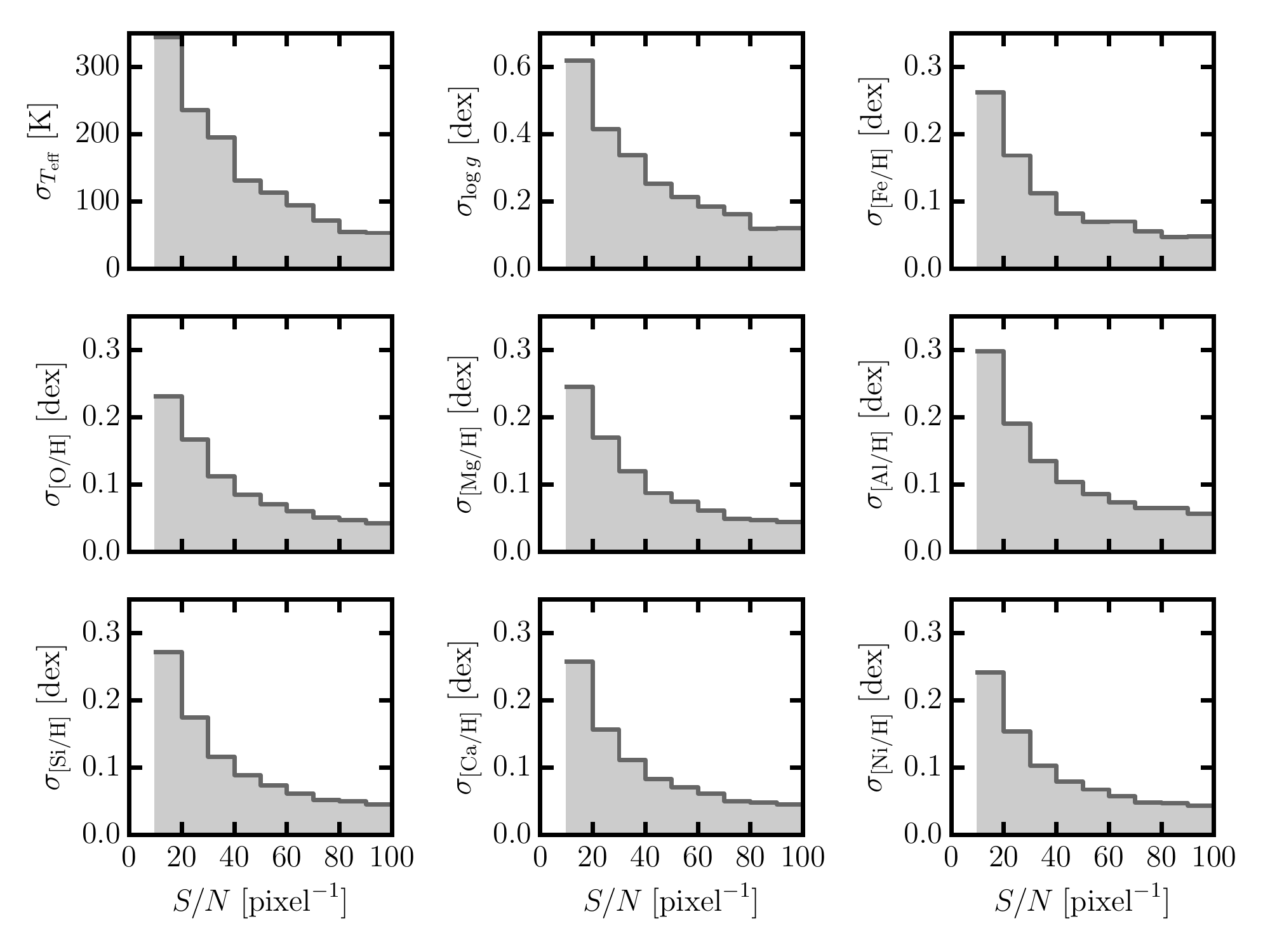}
\caption{The RMS deviation of labels for repeated observations in the test set.  The RMS deviation is binned as a function of the S/N ratio of the individual visit spectra.  The
precision flattens out at S/N $\gtrsim 40$~pixel$^{-1}$, where our results become systematics-limited.
\label{fig:repeat-visits}}
\end{figure*}

In Figure \ref{fig:repeat-visits} we show the RMS difference
in labels between the stacked spectra and single visit, binned by the
S/N ratio of the individual visit spectrum.  Here we only show stars
where the \emph{stacked} spectrum had S/N $>100$~pixel$^{-1}$ to ensure
that our baseline comparisons were in a region where we are dominated
by systematic uncertainties.  The precision in all labels tends to
flatten out past S/N $> 40$~pixel$^{-1}$, and the precision at high S/N
ratios is comparable to the minimum error floors we adopted in Section
\ref{sec:repeat-observations}.  The median S/N of \rave\ spectra is
50~pixel$^{-1}$, at which point our abundance precision is about 
0.07~dex, varying a few tenths of a dex between different elements.

\subsection{External validation}
\label{sec:external-validation}

\subsubsection{Comparison with \rave\ DR4}
\label{sec:validation-kordopatis}

\begin{figure*}[p]
\includegraphics[width=\textwidth]{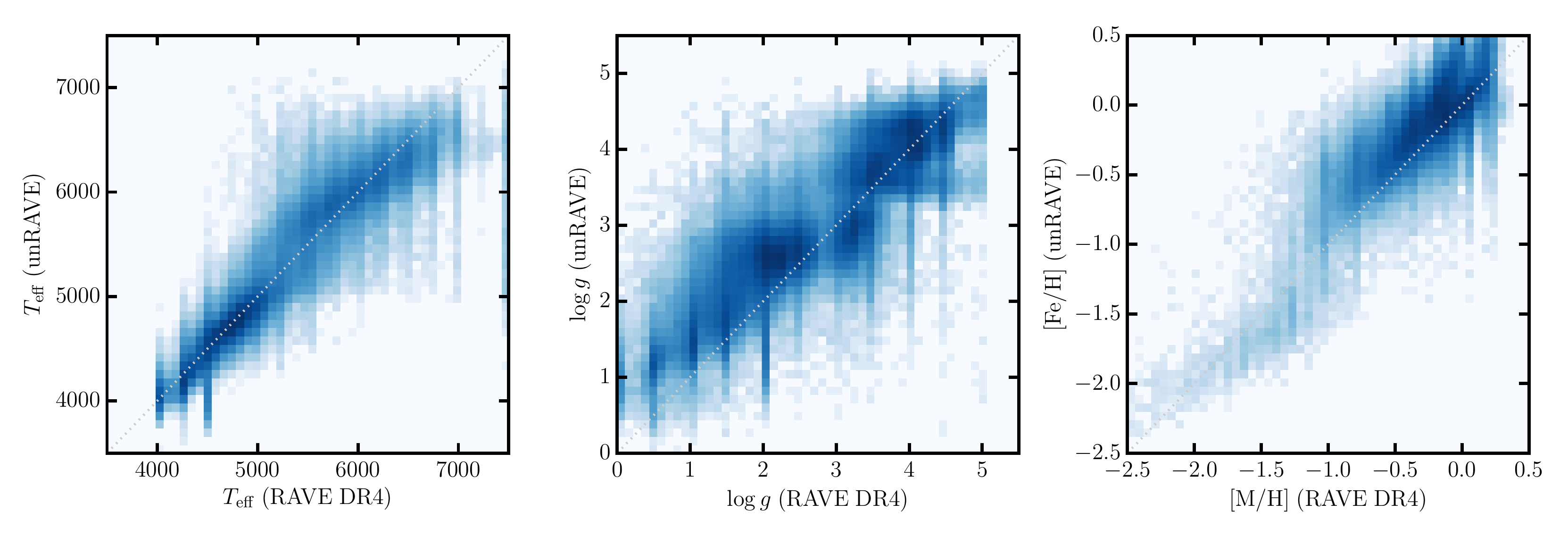}
\caption{Stellar parameter ($\teff$, $\logg$, [Fe/H]) comparison between the fourth \rave\ data release \citep{Kordopatis_2013} and this work. Here we show the `calibrated' metallicity (column \texttt{c\_M\_H\_K}) from the \rave\ survey. Only stars meeting quality constraints in \emph{both} studies are shown (see text for details).\label{fig:rave-dr4-comparison}}
\end{figure*}

We cross-matched our results against the official fourth \rave\ data release 
as an initial point of external comparison (Figure \ref{fig:rave-dr4-comparison}).
In order to provide a fair comparison, we only show stars that meet a number
of quality flags in \emph{both} samples.  Our constraints require that the
S/N ratio exceeds 10~pixel$^{-1}$, and $\chi_r^2 < 3$.  For this comparison
we further required that:
the \texttt{QK} flag from \citet{Kordopatis_2013} is zero, indicating no
problems were reported by the pipeline; $T_{{\rm eff},DR4} > 4000$~K;
the error in radial velocity \texttt{e\_HRV} is $<$8~km~s$^{-1}$; and the three
principal morphological flags \texttt{c1}, \texttt{c2}, \texttt{c3}, from 
\citet{Matijevic_2012} all indicate `n' for a normal FGK-type star.
There is good agreement in $\teff$, with a bias and RMS of just 4~K and 240~K,
respectively. The offset in $\logg$ on the giant branch between this study and 
\citet{Kordopatis_2013} has been noted in other studies (e.g., \apogee), and this issue
has been minimized in the fifth \rave\ data release by correcting $\logg$
values with a calibration sample consisting of asteroseismic targets and the
\gaia\ benchmark stars.  There is also a
slight discrepancy in the $\logg$ values along the main-sequence, where our
work tends to taper down towards higher $\logg$ values at cooler temperatures,
and the \rave\ DR4 sample tends to have a slightly flatter lower main-sequence.
This difference is not likely to have a very significant effect on the detailed
abundance or spectrophotometric distance determinations between these studies 
\citep{Binney_2014}.

\subsubsection{Comparisons with Reddy, Bensby, and Valenti \& Fischer}
\label{sec:validation-gold-standards}

We searched the literature for studies that overlap with \rave, and which base
their analysis on high-resolution, high S/N spectra.  We found four notable
studies with a sufficient level of overlap: the Milky Way disk studies by 
\citet{Reddy_2003,Reddy_2006} and \citet{Bensby_2014}, as well as the 
\citet{Valenti_Fischer_2005} work on exoplanet host star candidates.  These 
studies perform a careful (manual; expert) analysis using extremely 
high-resolution, high S/N spectra, and make use of \project{Hipparcos} 
parallaxes where possible.  Most of the stars in these samples are 
main-sequence or sub-giant stars.  Therefore, these works constitute an 
excellent comparison to evaluate the accuracy of our results on the 
main-sequence and sub-giant branch.

\begin{figure*}[p]
\includegraphics[width=\textwidth]{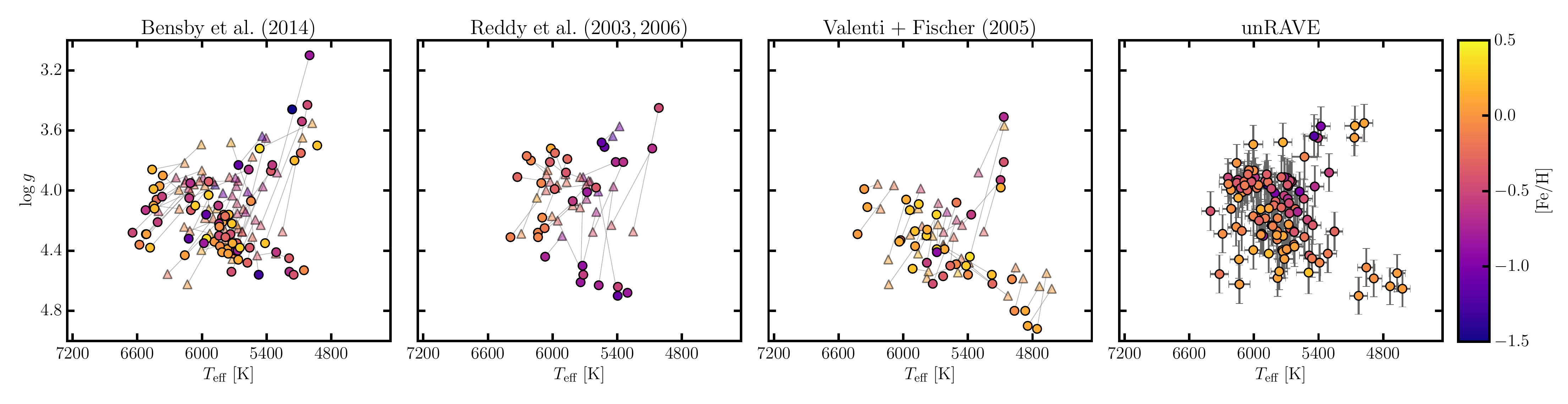}
\caption{Hertsprung-Russell diagrams of stars in common between this work and that of \citet{Bensby_2014,Reddy_2003,Reddy_2006,Valenti_Fischer_2005}.  Stars are colored by the metallicity of each study. Circles indicate literature markers in the first three panels, and the linked triangles indicate \raveon\ parameters for the same object. This figure illustrates the good qualitative agreement in the shape of the turnoff and sub-giant branch.\label{fig:gold-standard-hrd}}
\end{figure*}

\begin{figure*}[p]
\includegraphics[width=\textwidth]{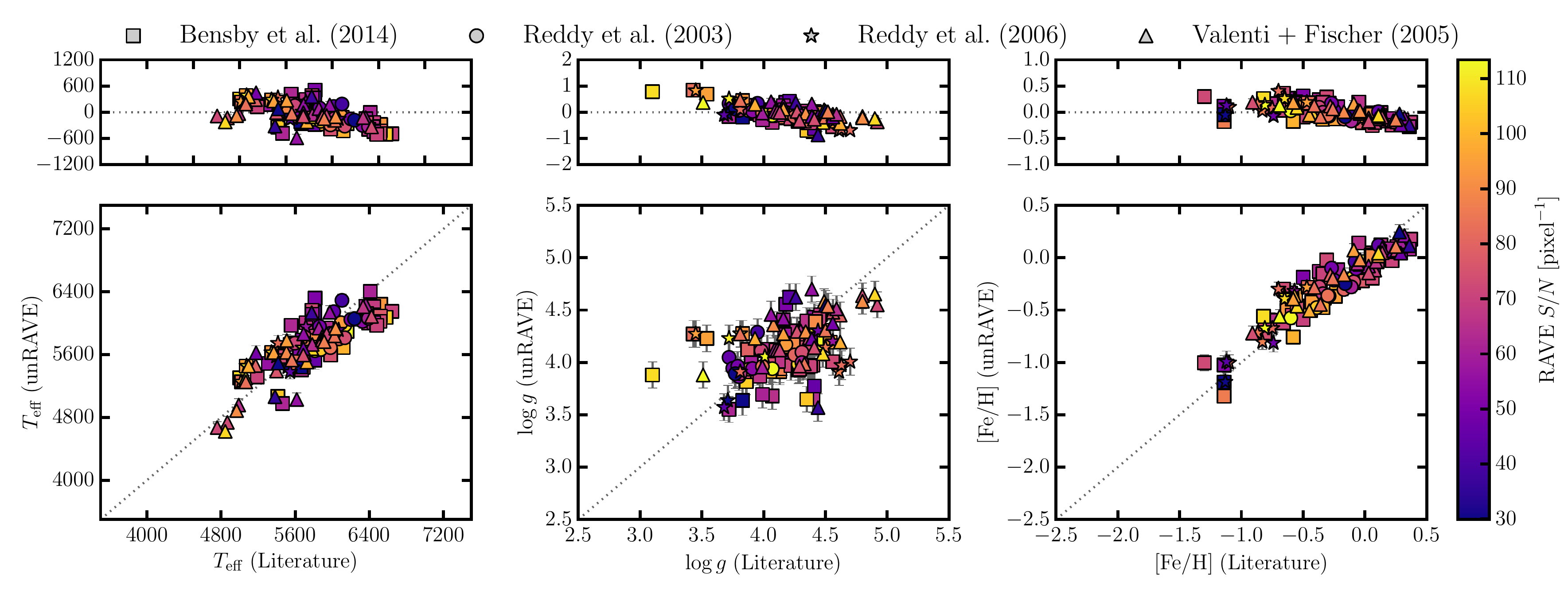}
\caption{Stellar parameter ($\teff$, $\logg$, [Fe/H]) comparisons for stars in common between this work and `gold standard' studies that use high-resolution, high S/N spectra and \hipparcos\ parallaxes where available: \citet{Bensby_2014,Reddy_2003,Reddy_2006,Valenti_Fischer_2005}. Stars are colored by the S/N of the \rave\ spectra. \label{fig:gold-standard-comparison}}
\end{figure*}

In Figure \ref{fig:gold-standard-hrd} we show Hertzsprung-Russell diagrams 
for the \rave\ stars that overlap with these studies.  We only include stars 
with $\chi_r^2 < 3$ and $S/N > 10$~pixel$^{-1}$, although the latter cut 
removed only a few stars because the average S/N in the \rave\ spectra for
these stars is relatively high ($\gtrsim{}50$~pixel$^{-1}$).  The literature 
data points in Figure \ref{fig:gold-standard-hrd} are linked to our derived
labels for the same stars, illustrating good qualitative agreement across 
the turn-off and sub-giant branch in all studies.  If we treat all three 
studies as a single point of comparison, the bias between our work and these 
studies is $-89$~K in $\teff$, just $-0.06$~dex in $\logg$, and $-0.03$~dex 
in [Fe/H] (see Figure~\ref{fig:gold-standard-comparison}).  The RMS deviation
in labels is 237~K, 0.30~dex, and 0.15~dex, respectively.  When considering 
the relative information content available in \rave\ (945~pixels in the near 
infrared with $\mathcal{R} \approx 7{,}500$) compared to these literature 
studies that use \project{Hipparcos} parallaxes where possible, and base 
their inferences on spectra with resolving power $\mathcal{R}$ between 
40,000 to 110,000, and S/N ratios exceeding 150, we consider the agreement 
to be very satisfactory.  Indeed, given the metallicity precision available
in the \raveon\ catalog, these results will likely be useful for future 
studies based on exoplanet host star properties (e.g., 
\project{TESS}\footnote{At present, however, there are just $\approx$30 
stars in \rave\ that overlap with the compilations of exoplanet host star 
properties listed at \texttt{exoplanets.org} and \texttt{exoplanets.eu}.}).

\subsubsection{Comparison with the \ges\ survey}
\label{sec:validation-ges}

\begin{figure*}[p]
\includegraphics[width=\textwidth]{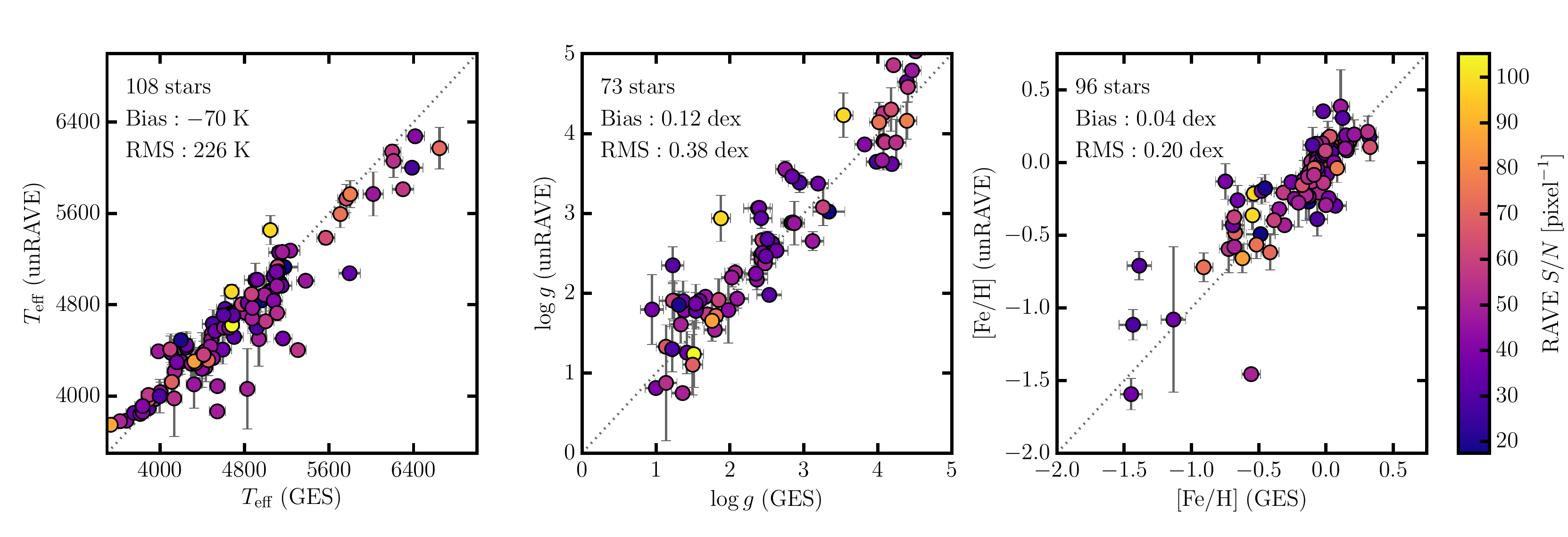}
\caption{Stellar parameter ($\teff$, $\logg$, [Fe/H]) comparison between the fourth internal data release from the \ges\ survey, and this work. The number of stars in each panel are shown, as well as the bias and RMS deviation in each label. Stars are colored by the S/N of the \rave\ spectra.  Most of the \ges/\rave\ overlap stars have relatively low S/N ratios in \rave, near $\approx 30$~pixel$^{-1}$.\label{fig:ges-stellar-parameters}}
\end{figure*}

There are 142 stars that overlap between \rave\ and the fourth internal
data release of the \ges\ survey. These are a mix of main-sequence, 
sub-giant and red giant branch stars.  About half (67) of the sample 
were acquired with the \acronym{UVES} instrument --- the other with the 
\acronym{GIRAFFE} spectrograph --- and the S/N of the \ges\ spectra peaks 
at $\approx140$~pixel$^{-1}$.  Despite most of these stars having relatively 
low S/N ratios in \rave\ ($\approx 25$~pixel$^{-1}$), there is good agreement
in with \ges\ and the \raveon\ stellar parameters (Figure~\ref{fig:ges-stellar-parameters}).  
The RMS in effective temperature, surface gravity and metallicity is 233~K, 
0.37~dex, and 0.17~dex, respectively.

Based on this comparison, we find no evidence for a systematic offset in
metallicities between stars on the main-sequence and those on the giant branch. 
This is a crucial observation, as the metallicities for stars in our 
main-sequence training set have a principally different source to those on 
the giant branch.  We cannot make these same inferences based on other 
surveys, like \apogee, because (1) \apogee\ stars formed part of the 
training set, and (2) they do not include main-sequence stars.  Even if we 
found good agreement between \epic\ and \apogee\ metallicities, this would 
not be informative, because \apogee\ is the source of metallicity for many 
stars on the giant branch in the \epic\ sample. Therefore, although this is
a qualitative comparison only, it is reassuring that there is no obvious 
systematic difference between the metallicities of main-sequence and giant
branch stars.

The metallicity agreement between this work and \ges\ extends down to low
metallicity, near $[{\rm Fe/H}] \approx -1.5$.  The scatter increases for
the few stars in the overlap sample with $[{\rm Fe/H}] < -1$, in the regime
where the influence from atomic transitions of these elements becomes very
small in \rave\ spectra. Moreover, these particular stars have lower S/N
ratios, which is reflected in the larger uncertainties
reported for these metallicities.

\begin{figure*}[p]
\includegraphics[width=\textwidth]{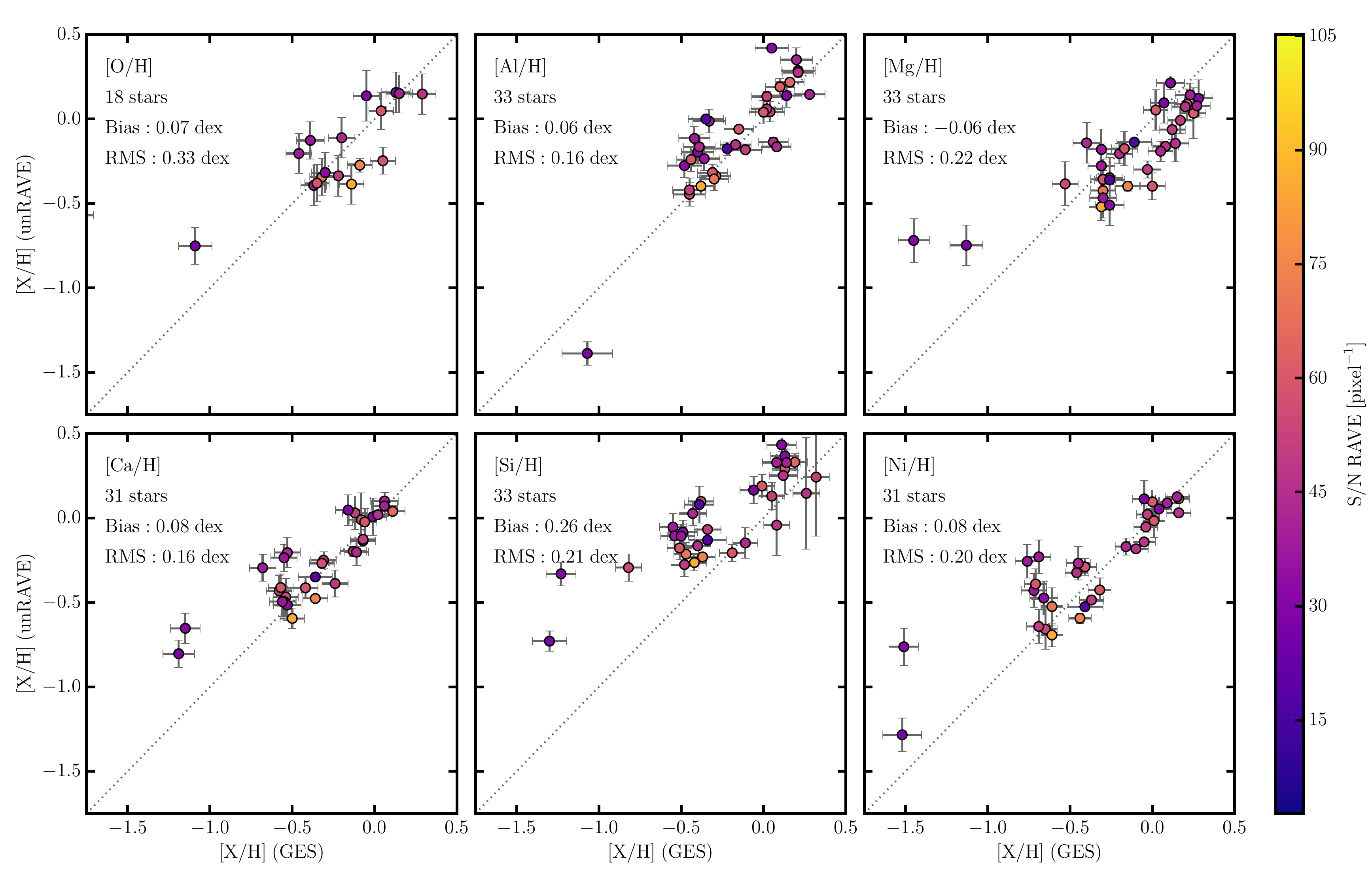}
\caption{Detailed chemical abundances in the fourth internal data release from the \ges\ survey compared to this work.  The number of stars shown in each panel is indicated, and the bias and RMS deviations are shown. Stars are colored by the S/N of the \rave\ spectra.\label{fig:ges-abundances}}
\end{figure*}

The fourth internal data release of the \ges\ includes detailed chemical
abundances of up to 45~species ($\approx32$ elements at different ionization
stages).  This provides us with an independent validation for our detailed
abundances on the giant branch.  These comparisons are shown in 
Figure~\ref{fig:ges-abundances}, where markers are colored by the
S/N of the \rave\ spectrum.  The number of stars available in each abundance
comparison varies due to what is available in the \ges\ data release, which
is itself a function of the instrument used, the spectral type, and other
factors.  The absolute bias for individual elements varies from as low as
0.06~dex ([Al,Mg/H]) to as high as 0.26~dex ([Si/H]), where we over-estimate 
[Si/H] abundances relative to the \ges\ survey.  The large bias in [Si/H]
is likely a consequence of an offset between [Si/H] abundances in \ges\ and
\apogee, the source of our training set for giant star abundances.  
The RMS deviation in each label is small for stars with [X/H] $> -0.5$, 
before increasing at lower metallicities.  If we consider all stars, the 
smallest abundance RMS we see with respect to \ges\ is 0.16~dex for [Ca/H] 
and [Al/H].  The increasing RMS at low metallicity is likely a consequence
of multiple factors, namely: inaccurate abundance labels for metal-poor stars
(Section~\ref{sec:the-training-set}); only weak, blended lines being available
in \rave, which cease to be visible in hot and/or metal-poor stars; and to a 
lesser extent, low S/N ratios for those particular stars being compared.  
Unfortunately not all of these factors are represented by the quoted errors 
in each label.  For these reasons, although it affects only a small number of
stars, we recommend caution when using individual abundances for very 
metal-poor giant stars in our sample.

\subsubsection{Comparison with the \rave\ DR4 calibration sample}
\label{sec:dr4-calibration-sample}

\begin{figure*}[p]
\includegraphics[width=\textwidth]{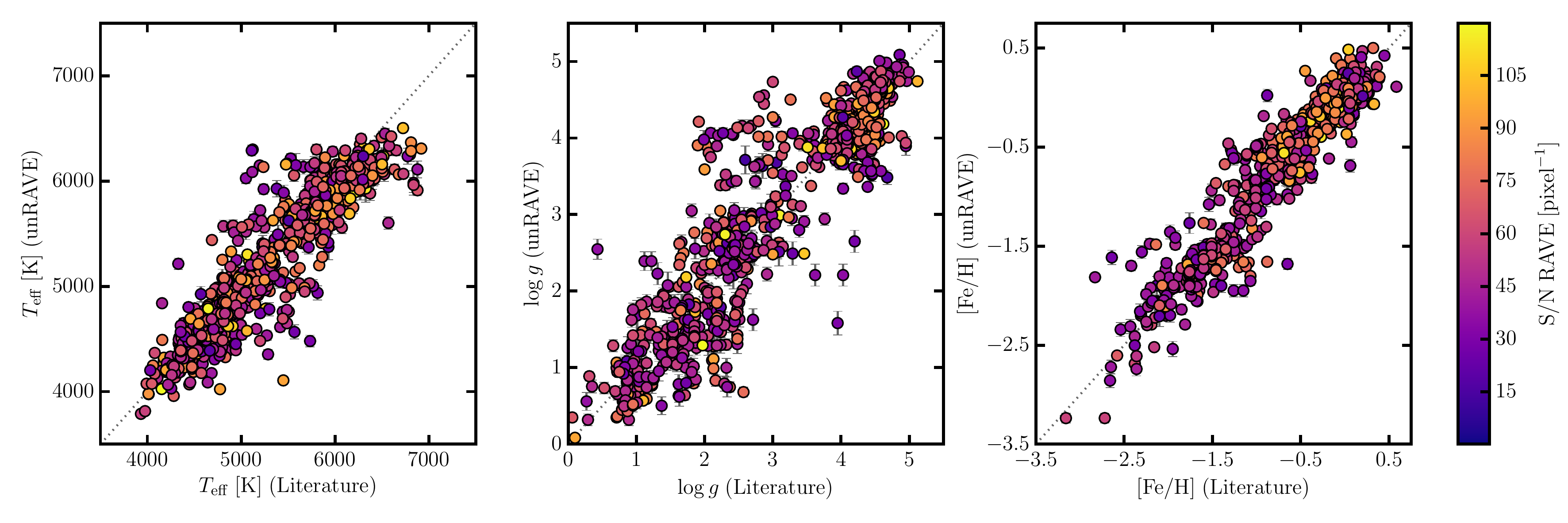}
\caption{Stellar parameter ($\teff$, $\logg$, [Fe/H]) comparison with the literature calibration sources used by \citet{Kordopatis_2013} and \citet{Kunder_2016}. Stars are colored by the S/N of the \rave\ spectra. Note that this comparison is for illustrative purposes only: it is not an indication of independent agreement with the literature because some metal-poor stars in this literature sample were used in the construction of our training set (see text for details).\label{fig:kordopatis-calibration}}
\end{figure*}

The fourth \rave\ data release made use of a number of high-resolution studies
to verify the accuracy of their derived stellar atmospheric parameters.  These
samples include main-sequence stars, giant stars, with a particular focus to
include metal-poor stars to identify (and correct) any deviations at low
metallicities.  We refer the reader to \citet{Kordopatis_2013} for the full 
compilation of literature sources.  Although the stellar atmospheric parameters
in this compilation come from multiple (heterogeneous) sources, we find 
generally good agreement with these works 
(Figure \ref{fig:kordopatis-calibration}).  However, we note that some 
reservation is warranted when evaluating this comparison, as some of the 
metal-poor stars in this calibration sample formed part of our training set.

\subsection{Astrophysical validation}
\label{sec:astrophysical-validation}

\subsubsection{Globular clusters}
\label{sec:globular-cluster-validation}

After verifying that our atmospheric parameters and abundances are comparable 
with high-resolution studies, here we verify that our results are consistent
with expectations from astrophysics.  In the \rave\ survey, \cite{Anguiano_2015}
identified 70 stars with positions and radial velocities that are consistent 
with being members of globular clusters: 49 stars belonging to NGC~5139
($\omega$~Centauri), 11 members of the retrograde globular cluster NGC~3201, and 10 
members of NGC~362. In addition, \citet{Kunder_2014} compiled 12 stars thought to 
belong to NGC~1851, and a further 10 stars in NGC~6752.  We refer the reader to 
those studies for details regarding the membership selection.

\begin{figure*}[p]
\center
\includegraphics[height=\textheight]{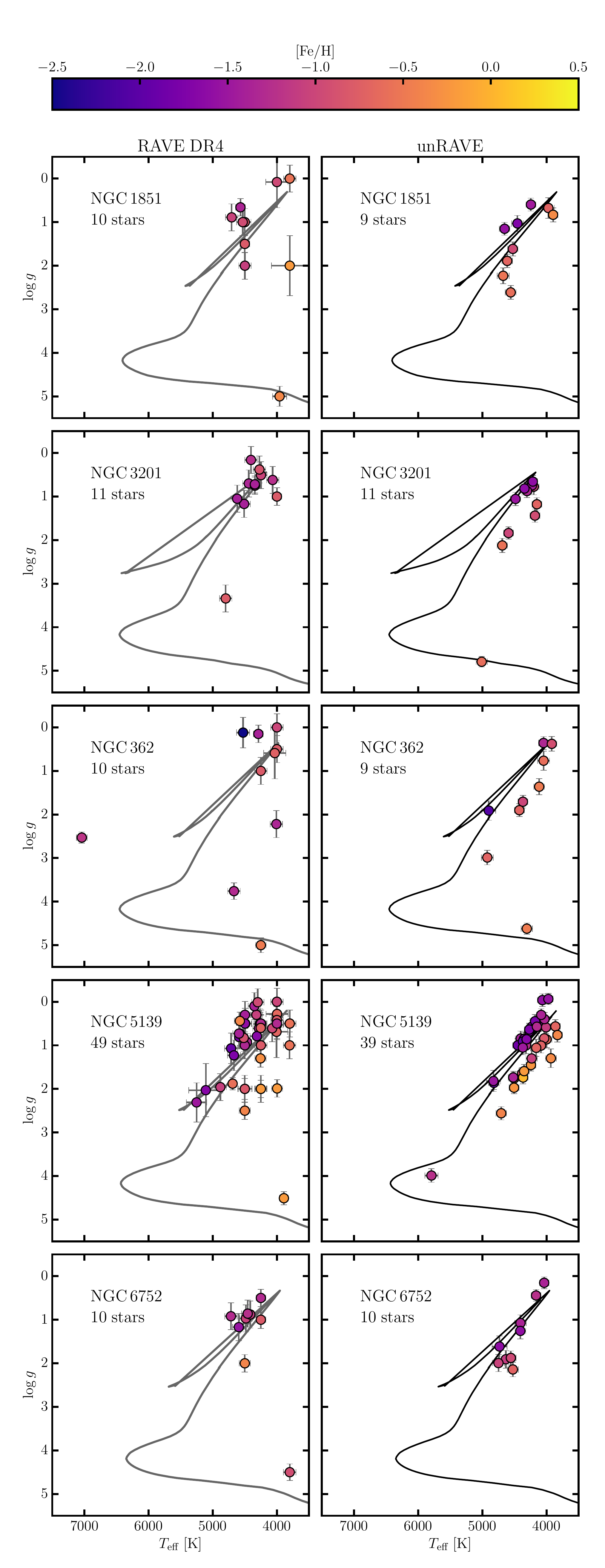}
\caption{Effective temperature $\teff$ and surface gravity $\logg$ for globular cluster members identified by \citet{Kunder_2014,Anguiano_2015}.  Left-hand panels indicate results from the fourth \rave\ data release \citep{Kordopatis_2013}, and the right-hand panels show results from this work.  A representative isochrone is shown for each cluster \citep{Bressan_2012}.\label{fig:globular-cluster-HRD}}
\end{figure*}

In Figure~\ref{fig:globular-cluster-HRD} we show our effective temperature
$\teff$ and surface gravity $\logg$ for these prescribed globular cluster members.
The right-hand panels indicate measurements made in this work, and for comparison 
purposes we have included the results from the fourth \rave\ data release in the 
left-hand panels.  We show representative \project{PARSEC} isochrones 
\citep{Bressan_2012} in all panels, where the isochrone ages and metallicities 
are adopted from \citet{Kunder_2016,Marin-Franch_2009}, and the 
\citet[][accessed 6 September 2016]{Harris_1996} catalog of globular 
cluster properties.  The globular cluster with the most number of members is 
NGC~5139 ($\omega$-Centauri), where we find a significant metallicity spread
that is consistent with high-resolution studies \citep{Marino_2011,Carretta_2009,
Carretta_2013}. Based on the pre-defined membership criteria, we find the mean 
metallicity of $\omega$-Centauri to be $[{\rm Fe/H}] = -0.85$.  However, it 
is clear that the membership criteria could be improved with our revised 
metallicities and detailed chemical abundances.  Indeed, our individual abundance
labels could be further used to identify globular cluster members --- at least, of
relatively metal-rich clusters ---  that are now tidally disrupted 
\citep{Anguiano_2016,Kuzma_2016,Navin_2016}, even for stars with low S/N ratios.

\subsubsection{Open clusters}
\label{sec:open-cluster-validation}

\begin{figure*}[p]
\center
\includegraphics[height=\textheight]{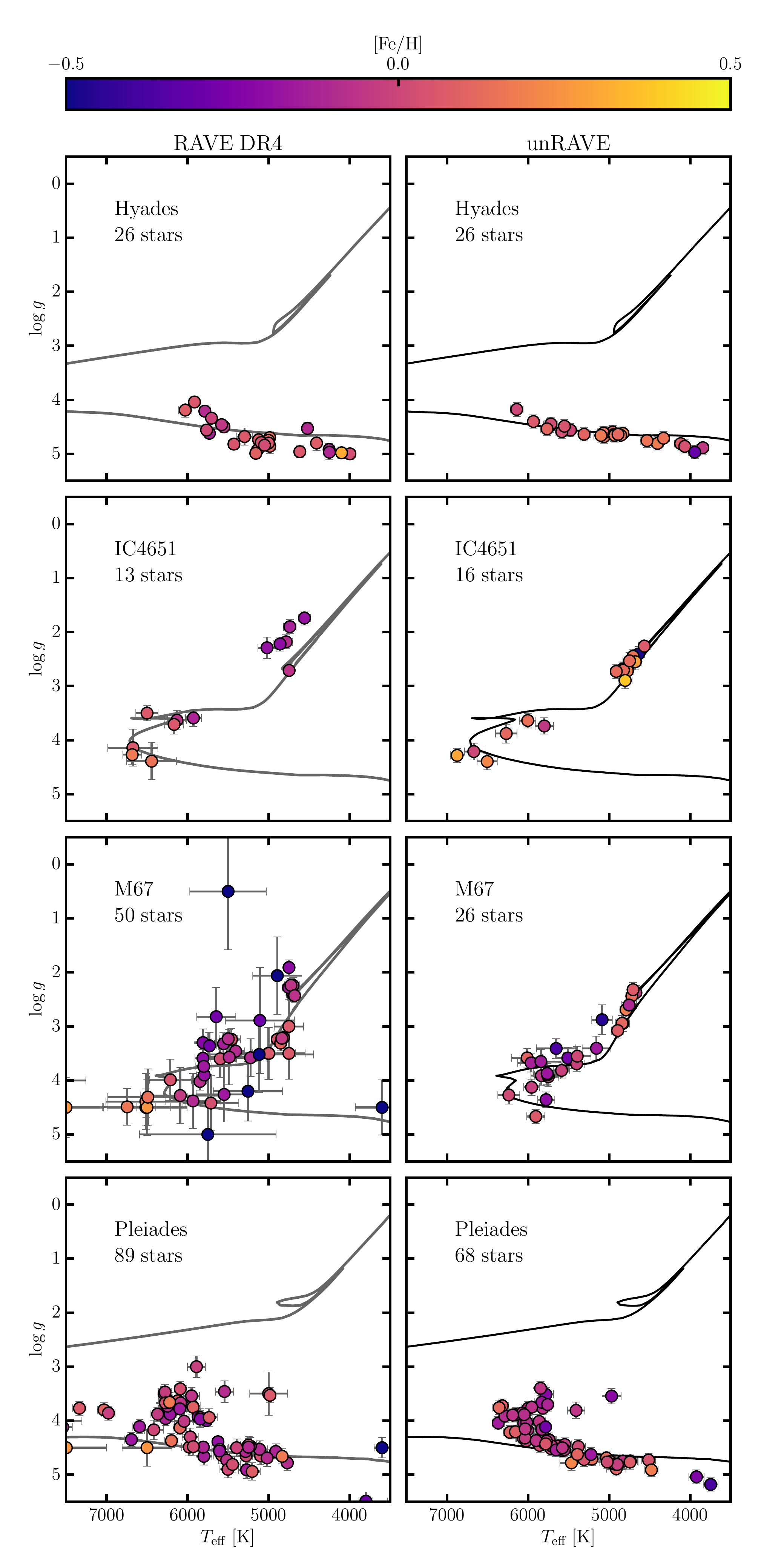}
\caption{Effective temperature $\teff$ and surface gravity $\logg$ for open cluster members identified by \citep{Kunder_2016}.  Left-hand panels indicate results from the fourth \rave\ data release \citep{Kordopatis_2013}, and the right-hand panels show results from this work.  A representative isochrone is shown for each cluster \citep{Bressan_2012}.\label{fig:open-cluster-HRD}}
\end{figure*}

Using positions, proper motions, and metallicities from the \rave\ survey
(i.e., not ours, such that they can be used as comparison), we identify $\sim$160 
probable members of four open clusters that were observed by \rave.
Specifically we identify 78 potential Pleiades members, 26 candidates in the
Hyades, another 13 in IC4561, and 30 stars in the solar-metallicity open
cluster M67.
We show the effective temperature $\teff$ and surface gravity $\logg$ for these
cluster candidates in Figure \ref{fig:open-cluster-HRD}.  The isochrones are 
sourced from \citet{Bressan_2012}, with cluster properties adopted from 
\cite{Kharchenko_2013}.

We find good agreement between our atmospheric parameters (right-hand panels) and the 
isochrones shown.  The position of the red clump in IC4651 and M67 are perfectly matched
to the isochrone, and the Hyades main-sequence is in good agreement down to 
$T_{\rm eff} \approx 4000$~K.  Similarly, we find consistency with the literature and
our metallicity scale.  We find the mean metallicity of M67 stars to be 
$[{\rm Fe/H}] = -0.02 \pm 0.03$~dex, in excellent agreement with the expected 
$[{\rm Fe/H}] = 0.00$ value (not accounting for atomic diffusion).
We further find the Hyades mean metallicity to be $[{\rm Fe/H}] = 0.07 \pm 0.09$~dex,
consistent with \citet{Paulson_2003}: $[{\rm Fe/H}] \approx 0.13$.
For IC4651 from 13 stars we find a mean $[{\rm Fe/H}] = 0.15 \pm 0.03$~dex, matching the
high-resolution, high S/N study of \citet{Pasquini_2004}, where they find 
$[{\rm Fe/H}] = 0.10 \pm 0.03$~dex. Finally, from 78 stars we find the mean
metallicity of the Pleiades to be $[{\rm Fe/H}] = -0.02 \pm 0.01$~dex, in very 
good agreement with the $[{\rm Fe/H}] = -0.034 \pm 0.024$~dex measurement reported by 
\citet{Friel_Boesgaard_1990}.

Despite the discrepancies between the isochrone and our derived labels for stars
in the Pleiades, we have made no attempt to refine the membership selection in
any of the aforementioned clusters.  We note however, that the same discrepancy
with the isochrone appears present in the fourth \rave\ data release \citet{Kordopatis_2013}.

\section{Discussion}
\label{sec:discussion}

We have performed an independent re-analysis of \Nspectra\ \rave\ spectra,
having derived atmospheric parameters ($\teff$, $\logg$, [Fe/H]) for all stars,
as well as detailed chemical abundances for red giant branch stars.  When 
combined with the \tgas\ sample, these results amount to a powerful compendium 
for chemo-dynamic studies of the Milky Way.  However, our analysis has caveats.
Inferences based on these results should recognize those caveats, and acknowledge 
that these results are subject to our explicit assumptions, some of which are provably
incorrect.

For practical purposes we adopted separate models: one for the giant branch and one
for the main-sequence.  A third model was used to derive relative weights for which
results to use.  The relative weighting we have used does not have any formal
interpretation as a likelihood or belief (in any sense): it was introduced for
practical reasons to identify systematic errors and combine results for multiple
models.  Because the relative weights have no formal interpretation, it is reasonable
to consider this method is as \emph{ad hoc} as any other approach.  The relative 
weighting has no warranty to be (formally) correct, and therefore may introduce 
inconsistencies or systematic errors rather than minimizing them.

If we only consider the results from individual models, there are a number of cautionary
remarks that stem from the construction of the training set.  The labels for red giant 
branch stars primarily come from \apogee, where previous successes with \thecannon\
have demonstrated that \apogee\ labels based on high S/N data can be of high fidelity
\citep{Ness_2015,Ness_2016,Ho_2016,Casey_2016b}.
However, the lack of metal-poor stars in the \apogee/\rave\ overlap sample produced
a tapering-off in the test set --- where \emph{no} stars had reliably reported metallicities
below that of the training set --- which forced us to construct a heterogeneous training
set.  The metal-poor stars included in this sample are from high-resolution studies
\citep{Fulbright_2010,Ruchti_2011}, but it is not known if the stellar parameters are
of high fidelity because we have a limited number of quality statistics available. 
Moreover, there is no guarantee that the stellar parameters \emph{or} abundances are 
on the same scale as \apogee\ \citep[and good reasons to believe they will not be; see][]{Smiljanic_2014}.

If the metallicities of metal-poor giant stars were on the same scale as the \apogee\
abundances, there is a larger issue in verifying that the main-sequence metallicities
and giant branch metallicities are on the same scale.  The training set for the
main-sequence stars includes metallicities from a variety of sources, including
\lamost, and the fourth \rave\ data release.  Even on expectation value, there is 
no straightforward manner to ensure that the main-sequence model and the red giant
branch model produce metallicities on the same scale.  We see no systematic offset
in metallicities of dwarf and giant stars that overlap between \rave\ and the \ges\
survey, suggesting that if there is a systematic offset, it must be small. 
Nevertheless, these are only verification checks based on $<1$\% of the data, and 
there is currently insufficient data for us to prove both models are on the same 
abundance scale.

For some of the most metal-poor giant stars in \rave\, we \emph{know} the abundances 
are not on the same scale as \apogee, because we were forced to adopt abundances for 
specific elements when they were unavailable.  Although we sought to adopt mean
level of Galactic chemical enrichment at a given overall metallicity, this is not a
representative abundance. Even if that is the \emph{mean} enrichment at that Galactic
metallicity, there is no requirement for zero abundance spread.  More fundamentally, 
we are incorrectly asserting that the element \emph{must} be detectable in the 
photosphere of the star.  There may be no transition that is detectable in that 
star, even with zero noise, because it is too weak to have any effect on the 
spectrum.  In the most optimistic case, this could be considered to be forcing the
model to make use of correlated information between abundances.  In a more 
representative (pessimistic) case, we are simply invoking what all abundances should 
be at low metallicity.

\begin{figure}[p]
\center
\includegraphics[width=0.65\textwidth]{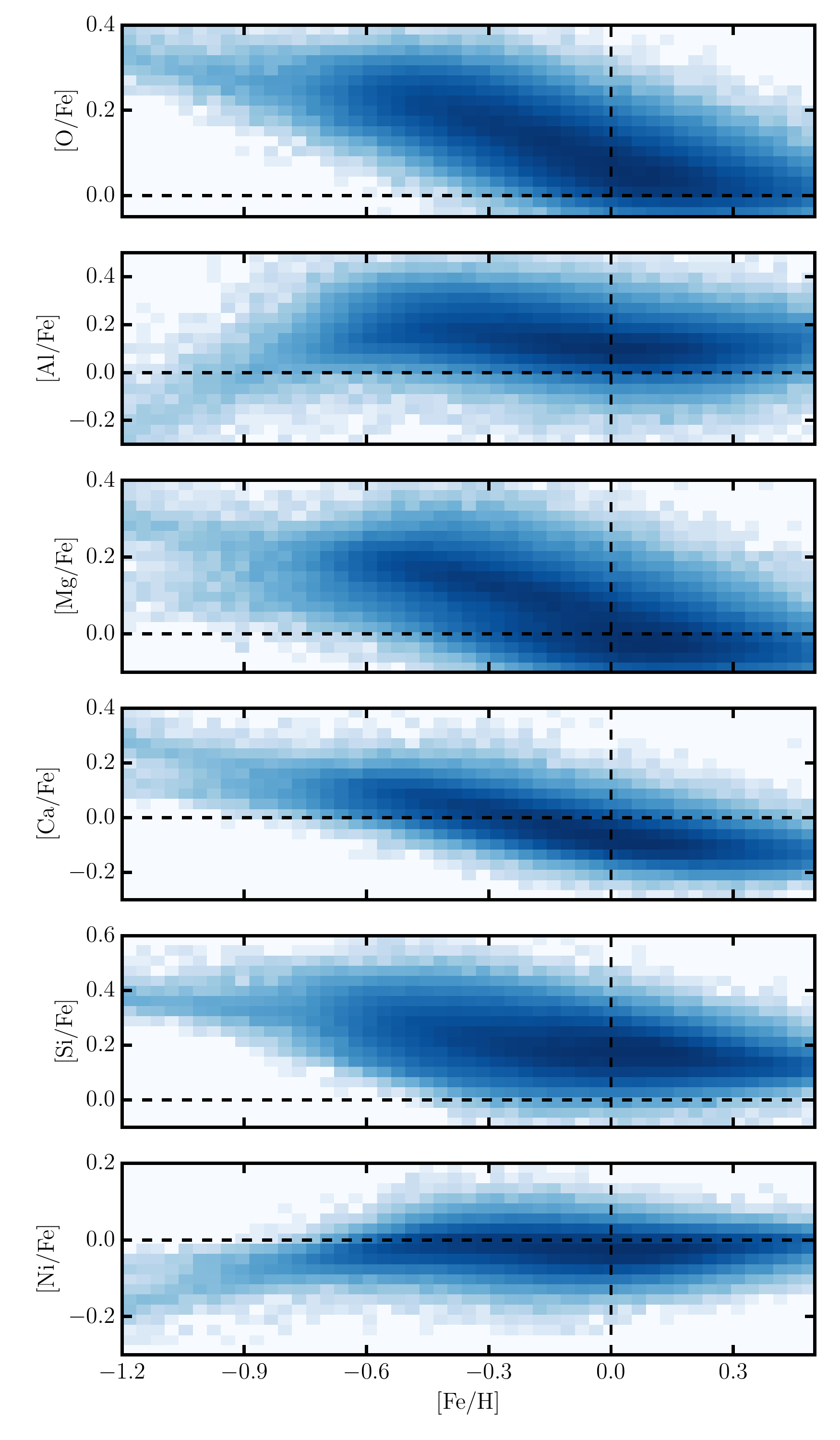}
\caption{Detailed chemical abundances ([X/Fe]) for giant stars in \raveon\ with respect to [Fe/H], showing the Galactic chemical evolution derived for each element. Bin densities are scaled logarithmically. Note that the y-axis limits vary for each panel, however for clarity we show the scaled-solar position by dashed lines, and have common tick mark spacing on the y-axis for all panels. Only stars meeting our quality constraints are shown (see Section \ref{sec:discussion}).\label{fig:gce}}
\end{figure}

This choice is reflected in the abundances of the test set.  While we do recover
trustworthy metallicities for ultra metal-poor ($[{\rm Fe/H}] \lesssim -4$) stars like 
CD-38~245, the individual abundances for all extremely metal-poor stars aggregate
(in [X/Fe] space; Figure~\ref{fig:gce}) at the assumed abundances for the metal-poor
stars in our sample.  Thus, while the overall metallicities appear reliable, the 
individual abundances for extremely metal-poor stars in the test set cannot be
considered trustworthy in any sense.  For this reason we have updated the electronic
catalog to discard these results as erroneous.

In Section~\ref{sec:method} we assumed that any fibre- or time-dependent variations
in the \rave\ spectra are negligible.  This is provably incorrect. Indeed,
\citet{Kordopatis_2013} note that the effective resolution of \rave\ spectra
varies from $6{,}500 < \mathcal{R} < 8{,}500$, and that the effective 
resolution is a function of temperature variations, fibre-to-fibre variations,
and thus position on the CCD \citep{Steinmetz_2006}. For this reason we ought
to expect our derived stellar parameters or abundances to be correlated either
with the fibre number, with the observation date, or both.  If significant, 
the trend could produce systematically offset stellar abundances solely due to
the fibre used.  \citet{Kordopatis_2013} conclude that resolution-based effects 
on the \rave\ stellar parameters should be a second-order effect. We have not 
seen evidence of these resolution-based correlations in our results, however, 
we have only performed cursory (non-exhaustive) experiments to investigate 
this issue.

We have shown some potential outcomes when the test set spectra differ 
significantly from the spectra in the training set.  Test set spectra that is
`unusual' from the training set can be projected as peculiar artefacts in 
label space.  In other words, unusual spectra can appear as `clumps' in
regions of parameter space that we could consider as being normal (e.g.,
an over-density of solar-type stars).  We addressed this issue for the main-sequence
and giant models by using a third model (Section \ref{sec:a-simple-model}) to
calculate relative weights.  However, spectra that are unusual from the 
training set used in the simple model could still project as systematic 
artefacts in label space.

\begin{figure}[p]
\center
\includegraphics[width=0.65\textwidth]{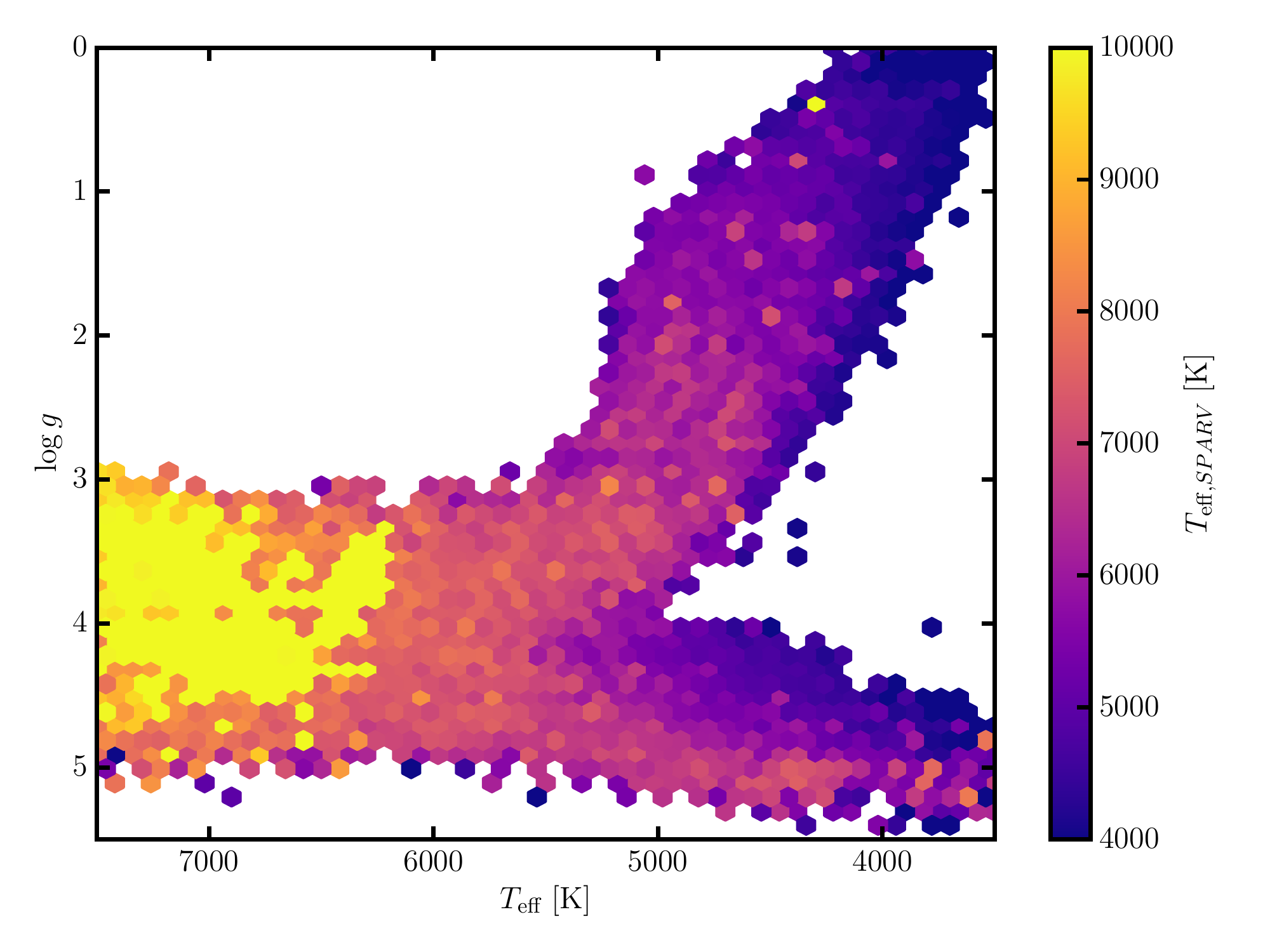}
\caption{Effective temperatures $\teff$ and surface gravities $\logg$ from this work, where each hexagonal bin is colored by the \emph{maximum} temperature for any star in that bin, as reported by the \rave\ pre-processing pipeline \texttt{SPARV} \citep{Steinmetz_2006,Zwitter_2008}.  Hot stars are not included in our training set, and appear above the turnoff in our labels.  For this reason, we supply a boolean quality control flag \texttt{QC} that flags stars (as \texttt{False}, for failing to meet our quality constraints) with $\teff > 8000$~K (as reported by \texttt{SPARV}), or results that do not meet other minimum constraints (see Section \ref{sec:discussion} for details).\label{fig:hot-stars}}
\end{figure}

Indeed, there are two known artefacts in our data that are relevant to this 
discussion. The first is a small over-density at the base of the giant branch, 
which is likely a consequence of joining the 9-label and 3-label models.  The
second has an astrophysical origin: there are no hot stars ($\teff > 8000$~K)
present in our training set, yet there are many in the \rave\ survey.
However, the \rave\ pre-processing pipeline 
\citep[\texttt{SPARV};][]{Steinmetz_2006,Zwitter_2008} performs template 
matching against grids of cool \emph{and} hot stars, and therefore we can use 
that information to identify hot stars.  In Figure \ref{fig:hot-stars} 
we show our derived effective temperatures $\teff$ and surface gravities $\logg$,
where each hexagonal bin is colored by the \emph{maximum} temperature
reported by \texttt{SPARV} for any star in that bin.  We show the maximum
temperature reported by \texttt{SPARV} to demonstrate that hot stars project
into a single clump in our label space --- near the turn-off --- in a region
where we may otherwise be deceived into thinking the observed over-density
is consistent with expectations from astrophysics.

This line of reasoning extends to spectra with other peculiar characteristics 
(e.g., chromospheric emission), and for these reasons we recommend the use of
additional metadata to investigate possible artefacts.  In our catalog we have
included a column containing a boolean flag to indicate whether the labels pass
very weak quality constraints.  Specifically, we flag results as failing our
quality constraints if \texttt{SPARV} indicates a $\teff > 8000$~K, or if 
$\chi_r^2 > 3$, or if $S/N < 10$~pixel$^{-1}$.  These quality constraints 
represent the minimum acceptable conditions and should not be taken verbatim: 
judicious use of the morphological classifications \citep{Matijevic_2012} or 
additional metadata from the \rave\ pre-processing pipelines is strongly 
encouraged.

We have presented a comprehensive collection of precise stellar abundances 
for stars in the first \gaia\ data release.  In total we derive stellar 
atmospheric parameters for \ReportedStellarParameters\ stars, and report more 
than 1.69 million abundances.  Despite the caveats and limitations discussed here, 
our validation experiments and comparisons with high-resolution spectroscopic
studies suggests that our results have sufficient accuracy and precision to
be useful for chemo-dynamic studies that become imminently feasible only in
the era of \gaia\ data.  We are optimistic that the \raveon\ catalog will 
advance understanding of the Milky Way's formation and evolution.

\subsection*{Access the results electronically}

\noindent{}Source code for this project is available at \texttt{\giturl},
and this document was compiled from revision hash \texttt{\githash} in that repository.
Derived labels, associated errors, and relevant metadata are available electronically
through the \rave\ database from 19 September 2016 onwards. Please note that it is a 
condition of using these results that the \rave\ data release by \citet{Kunder_2016} 
must also be cited, as the work presented here would not have been possible without the 
tireless efforts of the entire \rave\ collaboration, past and present.

\acknowledgements
We thank 
	Borja Anguiano (Macquarie),
	Jonathan Bird (Vanderbilt),
	Sven Buder (MPIA), 
	Sofia Randich (INAF),
and
	Kevin Schlaufman (Carnegie Observatories).
This research made use of: 
  	NASA's Astrophysics Data System Bibliographic Services;
  	Astropy, a community-developed core Python package for Astronomy \citep{astropy};
and 
  	\project{TOPCAT} \citep{Taylor_2005}.
This work was partly supported by the European Union FP7 programme through ERC 
grant number 320360. A.~R.~C. thanks King's College of the University of Cambridge
for their support.  Funding for K.~H.~ has been provided through the Simons 
Foundation Society of Fellows and the Marshall Scholarship.
Funding for RAVE has been provided by: the Australian Astronomical Observatory; 
the Leibniz-Institut fuer Astrophysik Potsdam (AIP); the Australian National 
University; the Australian Research Council; the French National Research Agency;
the German Research Foundation (SPP 1177 and SFB 881); the European Research 
Council (ERC-StG 240271 Galactica); the Istituto Nazionale di Astrofisica at 
Padova; The Johns Hopkins University; the National Science Foundation of the USA
(AST-0908326); the W. M. Keck foundation; the Macquarie University; the 
Netherlands Research School for Astronomy; the Natural Sciences and Engineering 
Research Council of Canada; the Slovenian Research Agency; the Swiss National 
Science Foundation; the Science \& Technology Facilities Council of the UK; 
Opticon; Strasbourg Observatory; and the Universities of Groningen, Heidelberg 
and Sydney. The RAVE web site is https://www.rave-survey.org.

\clearpage

\clearpage

\noindent{}$^{1}${Institute of Astronomy, University of Cambridge, Madingley Road, Cambridge CB3 0HA, UK} \\
$^{2}${Simons Center for Data Analysis, 160 Fifth Avenue, 7th floor, New York, NY 10010, USA} \\
$^{3}${Center for Cosmology and Particle Physics, Department of Physics, New York University, 4 Washington Pl., room 424, New York, NY 10003, USA} \\
$^{4}${Center for Data Science, New York University, 726 Broadway, 7th floor, New York, NY 10003, USA} \\
$^{5}${Max-Planck-Institut f\"ur Astronomie, K\"onigstuhl 17, D-69117 Heidelberg, Germany} \\
$^{6}${University of Ljubljana, Faculty of Mathematics and Physics, Jadranska 19, 1000 Ljubljana, Slovenia} \\
$^{7}${Research School of Astronomy and Astrophysics, Mount Stromlo Observatory, The Australian National University, ACT 2611, Australia} \\
$^{8}${Mullard Space Science Laboratory, University College London, Holmbury St Mary, Dorking, RH5 6NT, UK} \\
$^{9}${Observatoire astronomique de Strasbourg, Universit\'e de Strasbourg, CNRS, UMR 7550, 11 rue de l'Universit\'e, F-67000 Strasbourg, France} \\
$^{10}${Sydney Institute for Astronomy, School of Physics, University of Sydney, NSW 2006, Australia} \\
$^{11}${E.A. Milne Centre for Astrophysics, University of Hull, Hull, HU6 7RX, United Kingdom} \\
$^{12}${Astronomisches Rechen-Institut, Zentrum f\"ur Astronomie der Universit\"at Heidelberg, M\"onchhofstr.\ 12--14, 69120 Heidelberg, Germany} \\
$^{13}${Kapteyn Astronomical Institute, University of Groningen, P.O. Box 800, 9700 AV Groningen, The Netherlands} \\
$^{14}${INAF Astronomical Observatory of Padova, 36012 Asiago (VI), Italy} \\
$^{15}${Department of Physics and Astronomy, University of Victoria, Victoria, BC, Canada V8P 5C2} \\
$^{16}${Senior CIfAR Fellow} \\
$^{17}${Department of Physics and Astronomy, Macquarie University, Sydney, NSW 2109, Australia} \\
$^{18}${Western Sydney University, Penrith South DC, NSW 1797} \\
$^{19}${Johns Hopkins University, Baltimore, MD, USA} \\

\end{document}